\documentclass[a4paper,11pt]{article}
\pdfoutput=1 
\usepackage{jheppub} 

\usepackage[T1]{fontenc} 

\RequirePackage[displaymath]{lineno} 
\usepackage{epstopdf}
\usepackage{epsfig}
\usepackage{graphicx}
\usepackage{dcolumn}
\usepackage{bm}
\usepackage{ltablex,booktabs}
\usepackage{overpic}
\usepackage{subfigure}
\usepackage{float}
\usepackage{color}
\usepackage{amsmath}
\usepackage{mathcomp}
\usepackage{mathrsfs}
\usepackage{multirow}
\usepackage{makecell}
\usepackage{rotating}
\usepackage{amssymb}
\usepackage{gensymb}
\usepackage{amsmath}
\usepackage{tabularx}
\usepackage{threeparttable}

\title{Amplitude analysis of $\psi(3686)\to \gamma K_S^0  K_S^0 $}

\collaborationImg{\includegraphics[width=0.15\textwidth, angle=90]{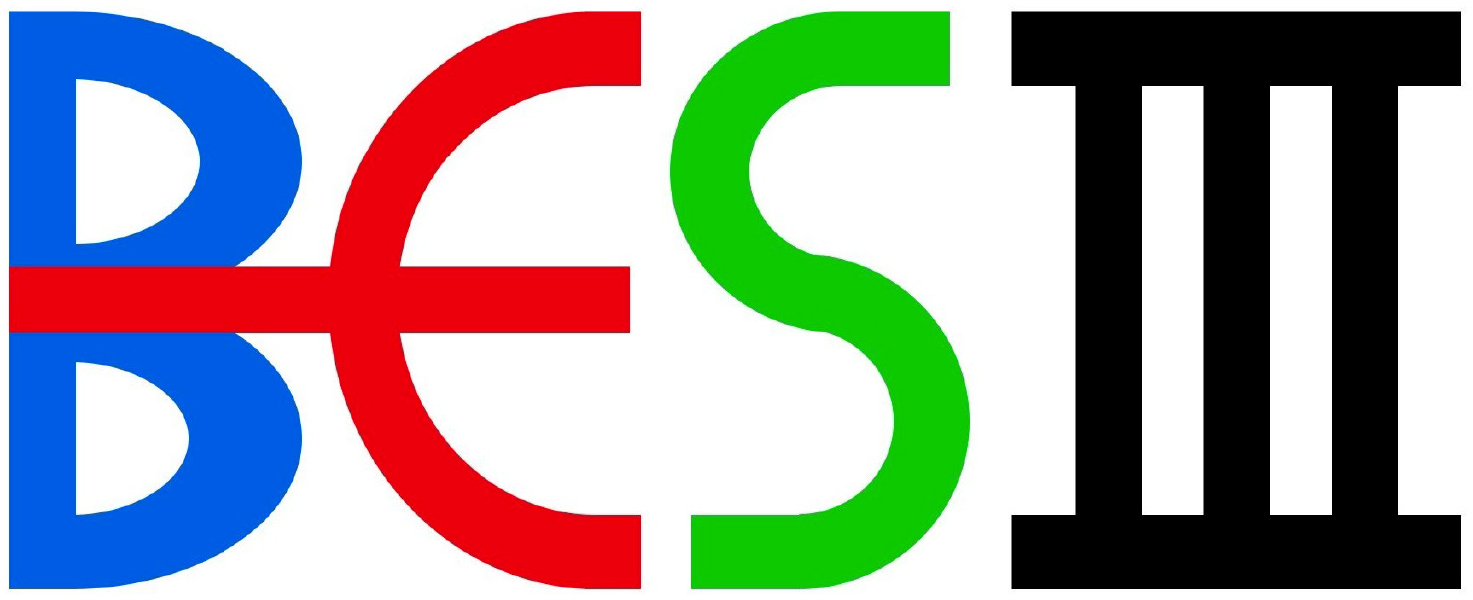}}
\collaboration{BESIII Collaboration}

\abstract{Using $(2712\pm14)\times10^6$ $\psi(3686)$ events collected with the BESIII detector, we perform the first amplitude analysis of the radiative decay $\psi(3686)\to \gamma K_S^0  K_S^0$ within the mass region $M_{K_S^0  K_S^0 }<2.8$ GeV/$c^2$. Employing a one-channel K-matrix approach for the description of the dynamics of the $K^0_S K^0_S$ system, the data sample is well described with four poles for the $f_0$-wave and three poles for the $f_2$-wave. The determined pole positions are consistent with those of well-established resonance states. The observed $f_0$ and $f_{2}$ states are found to be in agreement with those produced in radiative $J/\psi$ decays. The production behaviors of $f_0$ and $f_2$ poles in $\psi(3686)\to\gamma K_S^0 K_S^0$ are qualified with their residues and the converted branching fractions. By comparing with $J/\psi\to\gamma K_S^0 K_S^0$ decay, the ratios $\frac{\mathcal{B}(\psi(3686)\to\gamma f_{0,2})}{\mathcal{B}(J/\psi\to\gamma f_{0,2})}$ are determined, which provides crucial experimental inputs on the internal structure of the $f_{0,2}$ states, especially their potential mixing with glueball components.
}
\arxivnumber{}

\begin{document}
\maketitle
\flushbottom

\section{Introduction}

Quantum chromodynamics~(QCD) is now widely accepted as the theory for the strong interaction. In the framework of perturbative QCD, remarkable successes have been achieved in the prediction of hadronic phenomena in the high energy region such as heavy quarkonia~\cite{Brambilla:2005zw}. In the non-perturbative region, several theoretical investigations have been conducted on glueball states, including bag models~\cite{Jaffe:1975fd,Barnes:1981kq}, QCD-based potential models~\cite{Barnes:1981ac,Cornwall:1982zn,Brau:2004xw}, QCD sum rules~\cite{Shifman:1978by,Narison:1996fm}, and Lattice QCD~\cite{Bali:1993fb,Sexton:1995kd,Morningstar:1999rf,Loan:2005ff,Chen:2005mg,Richards:2010ck}, with experimental confirmation still lacking as of now. 
Most studies suggest that the lightest glueball state has a scalar quantum number $J^{PC}=0^{++}$ and a mass between $(1.45,1.75)$ GeV/$c^2$~\cite{Sexton:1995kd,Bali:2000vr,Chen:2005mg,Sun:2017ipk}. With these properties, glueballs are likely to mix with other mesons with the same quantum number, such as the $f_0$ states. This property makes it rather challenging to identify them among several conventional $q\bar{q}$ states, necessitating a detailed study of the $f_0$ spectrum.

Experimentally, the radiative decays of the $\psi$ states into two pseudo-scalar mesons via $\psi\to\gamma gg$, are particularly important for studying $f_0$ states and their potential mixing with the lightest scalar glueball. Conservation of parity and angular momentum restricts the quantum numbers to be $J^{PC}=\rm{even}^{++}$ for systems consisting of two identical pseudo-scalars. With an unprecedentedly large $J/\psi$ data sample, the BESIII collaboration has extensively studied the radiative decays of $J/\psi$ into $\pi^0\pi^0$~\cite{BESIII:2015rug}, $K_S^0  K_S^0 $~\cite{BESIII:2018ubj}, $\eta\eta$~\cite{BESIII:2013qqz}, $\eta^\prime\eta^\prime$~\cite{BESIII:2022zel}, and $\eta\eta^\prime$~\cite{BESIII:2022iwi,BESIII:2022riz}. These studies have revealed abundant $f_0$ and $f_2$ structures, and precise measurements of their properties have been performed. However, only few amplitude analyses of the radiative $\psi(3686)$ decays have been reported to date. For $\psi(3686)\to\gamma K_S^0  K_S^0 $, only a simple Breit-Wigner fit to the $M_{K_S^0  K_S^0 }$ spectrum has been performed based on the limited CLEO-c data set~\cite{Dobbs:2015dwa}. Extensive studies of $\psi(3686)$ radiative decays are therefore crucial to extract the $f_{0,2}$ poles and compare the productions of $f_{0,2}$ states between $J/\psi$ and $\psi(3686)$ radiative decays. 

The BESIII experiment has collected $(2712\pm14)\times10^6$ $\psi(3686)$ events~\cite{BESIII:2024lks}, which provide an excellent opportunity to investigate $f_0$ and $f_2$ states with the radiative decay $\psi(3686)\to \gamma K_S^0  K_S^0 $. In this article, we present the first amplitude analysis of $\psi(3686)\to \gamma K_S^0  K_S^0 $ with a one-channel K-matrix approach.

\section{Detector and data samples}
\label{sec:BES}

The BESIII detector~\cite{BESIII:2009fln} records symmetric $e^+e^-$ collisions provided by the BEPCII storage ring~\cite{Yu:2016cof} in the center-of-mass energy ($\sqrt{s}$) range from 1.85 to  4.95~GeV,
with a peak luminosity of $1.1 \times 10^{33}\;\text{cm}^{-2}\text{s}^{-1}$ achieved at $\sqrt{s} = 3.773\;\text{GeV}$. BESIII has collected large data samples in this energy region~\cite{BESIII:2020nme,Lu:2020bdc,Zhang:2022bdc}. The cylindrical core of the BESIII detector covers 93\% of the full solid angle and consists of a helium-based multilayer drift chamber~(MDC), a plastic scintillator time-of-flight system~(TOF), and a CsI(Tl) electromagnetic calorimeter~(EMC),
which are all enclosed in a superconducting solenoidal magnet providing a 1.0~T magnetic field. The solenoid is supported by an octagonal flux-return yoke with resistive plate counter muon
identification modules interleaved with steel.  The charged-particle momentum resolution at $1~{\rm GeV}/c$ is $0.5\%$, and the  ${\rm d}E/{\rm d}x$ resolution is $6\%$ for electrons from Bhabha scattering. The EMC measures photon energies with a resolution of $2.5\%$ ($5\%$) at $1$~GeV in the barrel (end cap) region. The time resolution in the TOF barrel region is 68~ps, while that in the end cap region was 110~ps. The end cap TOF system was upgraded in 2015 using multigap resistive plate chamber technology, providing a time resolution of 60~ps~\cite{Li:2017jpg,Guo:2017sjt,Cao:2020ibk}, which benefits 83\% of the data used in this analysis.


Simulated data samples produced with a {\sc geant4}-based~\cite{GEANT4:2002zbu} Monte Carlo (MC) package, which includes the geometric description of the BESIII detector and the detector response, are used to determine detection efficiencies and to estimate backgrounds. The simulation models the beam
energy spread and initial state radiation (ISR) in the $e^+e^-$ annihilations with the generator {\sc kkmc}~\cite{Jadach:1999vf}. The inclusive MC sample includes the production of the $\psi(3686)$ resonance, the ISR production of the $J/\psi$, and the continuum processes incorporated in {\sc kkmc}~\cite{Jadach:1999vf}. All particle decays are modeled with {\sc evtgen}~\cite{Lange:2001uf,Ping:2008zz} using branching fractions either taken from the particle data group (PDG)~\cite{ParticleDataGroup:2024cfk}, when available, or otherwise estimated with {\sc lundcharm}~\cite{Chen:2000tv,Yang:2014vra}. Final state radiation~(FSR) from charged final state particles is incorporated using the {\sc photos} package~\cite{Barberio:1990ms}.

\section{Event selection and background study}
\label{sec:selection}

Candidates for $\psi(3686) \to \gamma K_S^0  K_S^0 $ are reconstructed via the decay $K_S^0 \to \pi^+\pi^-$.
Charged tracks are required to originate from the region $|d_z|<20$ cm and $|\cos\theta|<$0.93, where $|d_z|$ is the distance of closest approach to the interaction point~(IP) along the $z$-axis, and $\theta$ is the polar angle relative to the $z$-axis, which is defined as the symmetry axis of the MDC. Only four charged tracks with zero net charge are allowed.

\begin{figure}[htbp]
	\centering
	\includegraphics[width=0.40\textwidth]{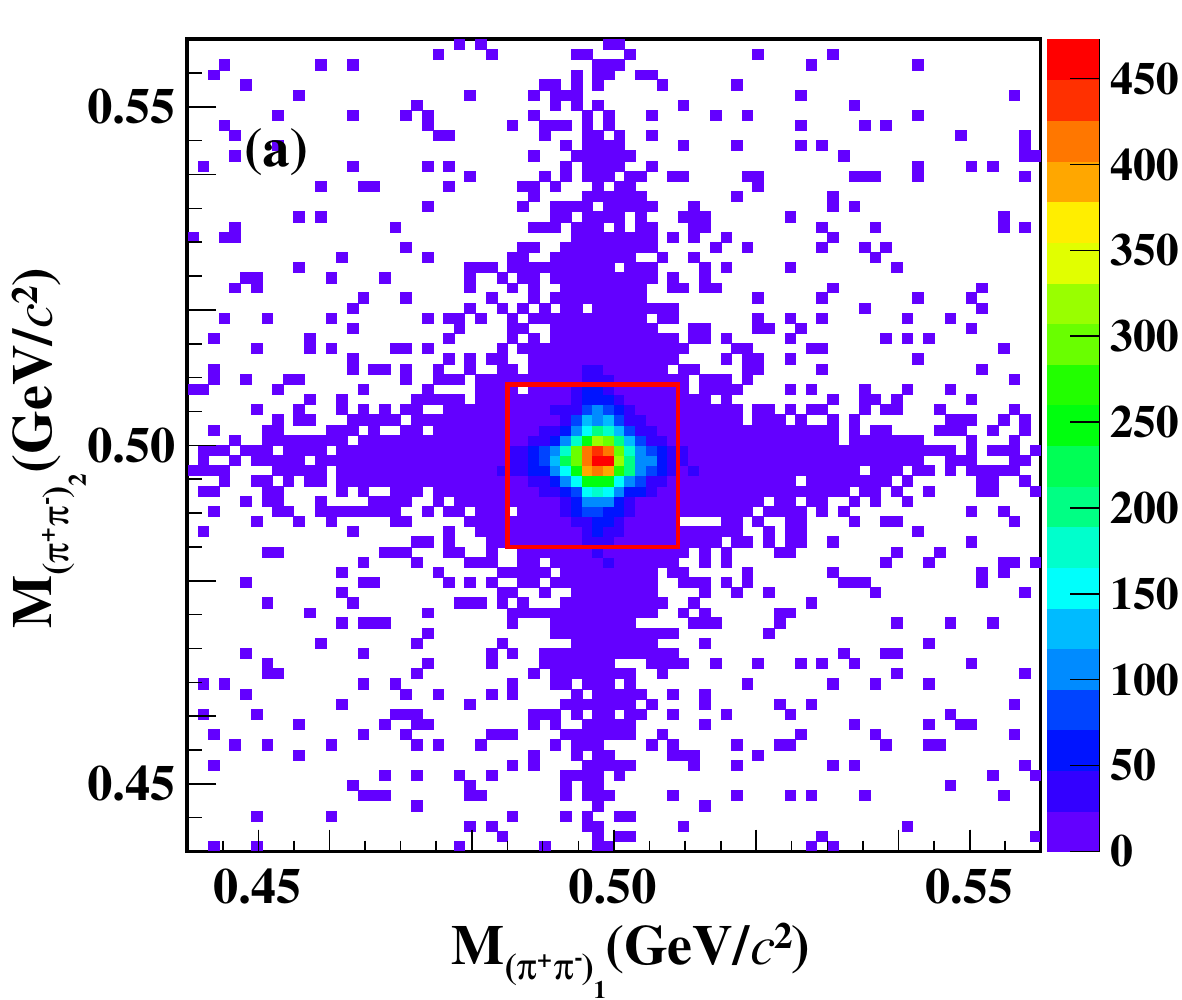}
	\includegraphics[width=0.54\textwidth]{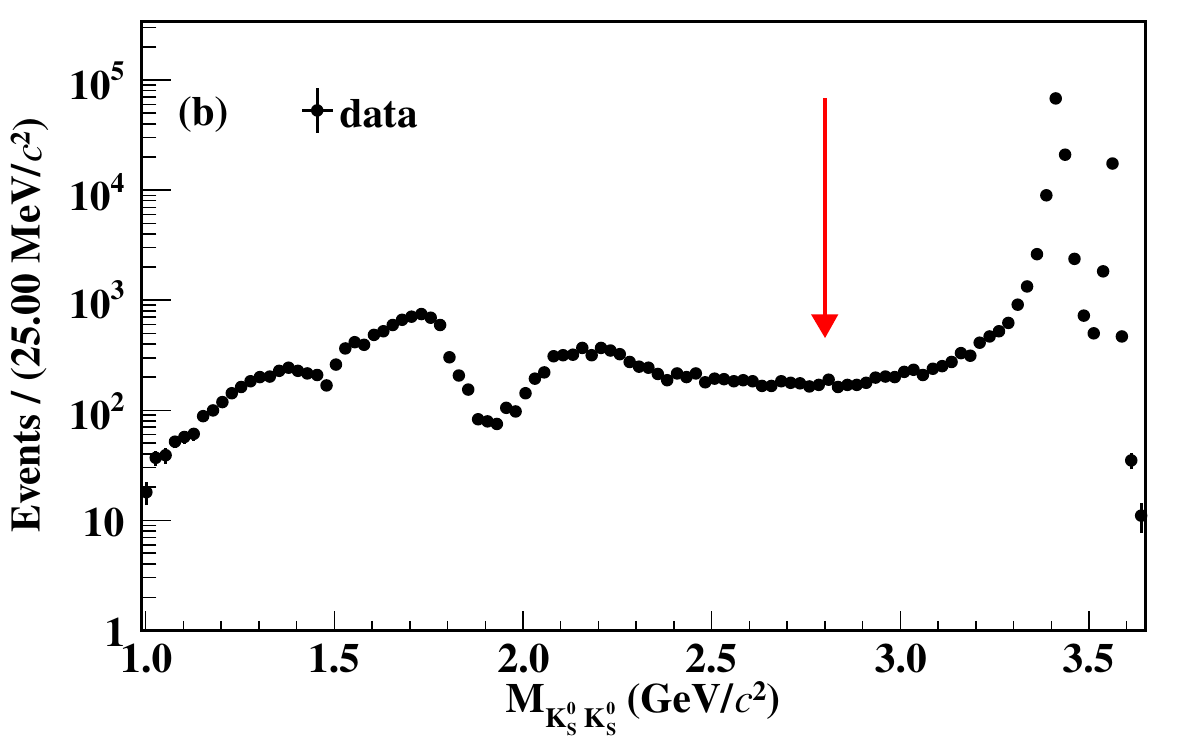}
	\caption{(a) The distribution of $M_{({\pi^+\pi^-})_1}$ versus $M_{({\pi^+\pi^-})_2}$ from the data sample. The red box delineates the $K_S^0 $ signal zone $|M_{\pi^+\pi^-}-M_{K_S^0 }|<12$ MeV/$c^2$. (b) The invariant mass of the $K_S^0 K_S^0$ system of the data sample in log-scale. The red arrow indicates the requirement $M_{K_S^0  K_S^0 }<2.8$ GeV/$c^2$.}
	\label{fig:KsKs_2D_plot}
\end{figure}

To reconstruct $K_S^0 $ candidates, all possible pairs of oppositely charged tracks satisfying the above requirement are assigned as $\pi^+\pi^-$ without particle identification. The $\pi^+\pi^-$  trajectories are constrained to originate from a common vertex by applying a vertex fit. The decay length of each $K_S^0 $ candidate, namely the distance between the IP and the decay vertex of $K_S^0 $, is required to be greater than twice its resolution. The invariant mass of $\pi^+\pi^-$ is required to satisfy $|M_{\pi^+\pi^-}-M_{K_S^0 }|<12$ MeV/$c^2$ corresponding to 2.5 times the resolution, where $M_{K_S^0 }$ is the known $K_S^0$ mass as quoted from PDG~\cite{ParticleDataGroup:2024cfk} and $M_{\pi^+\pi^-}$ is calculated at the IP for simplicity. The number of surviving $K_S^0 $ candidates is required to be exactly two. Figure~\ref{fig:KsKs_2D_plot}(a) shows the distribution of invariant masses of two $K_S^0 $ candidates, where a significant signal contribution is observed. In the subsequent analysis, two $K_S^0 $ candidates are mixed randomly as they are indistinguishable.

Photon candidates are identified using showers in the EMC. Each shower is required to have deposited at least 25 MeV in the barrel and 50 MeV in the endcaps of the EMC. 
To suppress contamination from charged particles, the angle between the shower position and the extrapolated trajectory of any charged track at the EMC must be greater than 10$^\circ$.
The timing difference between the EMC shower and the event start time is required to be within [0,~700] ns to suppress electronic noise and the energy deposits unrelated to the event. At least one good photon candidate is required.

To suppress the combinatorial background and improve the resolution, a four-constraint (4C) kinematic fit for the $\psi(3686)\to \gamma \pi^+ \pi^- \pi^+ \pi^- $ hypothesis is performed on the four-momenta of final-state particles in the lab frame. The combination with the smallest $\chi^2_{\rm 4C}$ is retained for further analysis, and the $\chi_{\rm 4C}^{2}$ of the kinematic fit is required to be less than 50. Contributions from $\chi_{c0,2}\to K_S^0  K_S^0$ are suppressed by requiring 
$M_{K_S^0  K_S^0 }<2.8$ GeV/$c^2$, as shown in Figure~\ref{fig:KsKs_2D_plot}(b).

A total of 17,672 candidates survived the above selection criteria. The background contribution is estimated with the $\psi(3686)$ inclusive MC sample, yielding a total background fraction from $\psi(3686)$ decays below 1\%, dominated by the final state $\gamma K_S^0 K_S^0 \pi^0$. This simulated background is incorporated into the further amplitude analysis. The background from the continuum process is estimated using a $(401.0\pm4.0)$~pb$^{-1}$ data sample at $\sqrt{s}=3.650$~GeV~\cite{BESIII:2024lks}, resulting in only one surviving event, which corresponds to 10 events in the $\psi(3686)$ data sample with integrated luminosity $(3877.0\pm39.0)$~pb$^{-1}$~\cite{BESIII:2024lks}. Consequently, the continuum background is ignored in the subsequent analysis.

\section{Amplitude analysis}
\label{sec:PWA}

\subsection{Amplitude}

The covariant tensor amplitude cited from Ref.~\cite{Zou:2002ar} is used to describe the radiative decay $\psi(3686) \to \gamma K_S^0 K^0_S$ in this work by assuming an isobar model. The amplitude is expressed as
\begin{equation}
\label{eq:amplitude}
\mathcal{A}=\psi_{\mu}(m_1)e^*_{\nu}(m_2) A^{\mu\nu} = \psi_{\mu}(m_1)e^*_{\nu}(m_2) \sum_{i} \Lambda_{i}U_{i}^{\mu\nu},
\end{equation}
where $\psi$ and $e$ denote the polarization vectors of $\psi(3686)$ and the photon, with polarizations $m_1$ and $m_2$, respectively. The complex coupling constant $\Lambda_{i}$ corresponds to the $i$-th amplitude component $U_{i}^{\mu\nu}$, whose explicit expressions are given in Ref.~\cite{Zou:2002ar}. Summing over the polarizations of $\psi(3686)$ and the photon, the amplitude squared is given by
\begin{equation}
\label{eq:cross_section}
	|\mathcal{M}|^2 = \frac{1}{2}\sum_{m_1=1}^{2} \sum_{m_2=1}^{2} |\mathcal{A}|^2.
\end{equation}
The Blatt-Weisskopf factor $B_L(Q,Q_0)$ has been incorporated into $U_{i}^{\mu\nu}$, where $Q$ is the momentum of daughter particles in the rest frame of their mother, and $Q_0 = \frac{0.197321}{R}$~GeV/$c$ is a hadron ``scale'' parameter. Here, $R$ denotes the radius of the centrifugal barrier in femtometers~(fm). In the nominal fit, we take $R=0.59$~fm, corresponding to 3 GeV$^{-1}$, which is the center of the expected range $[1, 5]$~GeV$^{-1}$~\cite{ParticleDataGroup:2024cfk}. Corresponding uncertainties are considered by varing $R$ in the systematic uncertainty.

The dynamic part $f^{(X)}$ for resonance $X$ has also been included in the $U_{i}^{\mu\nu}$ following the convention of Ref.~\cite{Zou:2002ar}. The $f_0$ and $f_2$ waves are described with the K-matrix approach. Since only one decay channel $f_{0,2}\to K_S^0  K_S^0 $ is considered in this work,
our methodology does not fully satisfy the unitarity condition since coupled-channel effects are neglected. Nonetheless, it offers a more dependable depiction of overlapping resonance states with significant widths compared to a straightforward summation of Breit-Wigner functions. To incorporate the production amplitude $\psi(3686)\to\gamma f_{0,2}$ within the K-matrix framework, a prevalent method called $\mathcal{P}$-vector parameterization~\cite{CrystalBarrel:2019zqh,Husken:2022yik} is used in this work. For the $i$-th production wave of $f_{0,2}$, corresponding dynamic part is written as
\begin{equation}
\label{eq:Fvector}
f^{(f_{0,2})}_{i}=n(1-\mathcal{K}i\rho n^2)^{-1}\mathcal{P}_{i},
\end{equation}
where $n=Q^L B_L(Q,Q_0)$ and $\rho= Q/\sqrt{s}$ is the phase space factor of $f_{0,2}\to K_S^0  K_S^0 $. 
The K-matrix  $\mathcal{K}$ is defined as $\mathcal{K}=\sum_{a}\frac{g^2_a}{m_a^2-s}+b$, where $m_a$ and $g_a$ are the bare mass and coupling to $K_S^0  K_S^0 $ for pole $a$, respectively, and $b$ accounts for the background contribution in the K-matrix. All parameters shown in $\mathcal{K}$ are real-valued. The production vector $\mathcal{P}$ is given as
\begin{equation}
\mathcal{P}_{i} = \sum_{a} \frac{\beta^{i}_{a} g_a}{m_a^2-s}+\beta^{i}_{\rm bkg},
\end{equation}
where the complex parameters $\beta^{i}_{a}$ describes the $i$-th wave production strength for pole $a$ and $\beta_{\rm bkg}$ models the non-resonant production. The $\beta$, $m$, and $g$ are all free parameters in the fit. Since the production strengths have already been included in the P-vector, the complex coupling constant $\Lambda_i$ in Eq.~\eqref{eq:amplitude} is fixed to one for both $f_0$ and $f_2$ components. The $K^*(892)$ resonance is described with a simple Breit-Wigner function $f^{(K^*)}=1/(M^2_0-s-i M_0 \Gamma_0)$, where the resonance parameters $M_0$ and $\Gamma_0$ are fixed to the ``charged only, hadroproduced'' values provided by the PDG. Considering the $K_S^0 K_S^0$ pair has been symmetrized, the complex coupling constants $\Lambda_i$ of the amplitudes of $K^*(892)\to K_{S1}^0\gamma$ and $K^*(892)\to K_{S2}^0\gamma$ are constrained to be the same in the fit.

\subsection{Fit method}
	The complex coupling constants $\Lambda$ and the parameters of the K-matrix shown in Eqs.~\eqref{eq:amplitude} and~\eqref{eq:Fvector} are determined by an event-based maximum likelihood fit. The log-likelihood function is constructed as
\begin{equation}
\ln \mathcal{L} = \ln \mathcal{L}_{\rm dt} - \ln \mathcal{L}_{\rm bg}.
\end{equation}
Here, $\ln \mathcal{L}_{\rm dt/bg}$ sums over all the data/background events and is defined as
\begin{equation}
\ln \mathcal{L}_{\rm dt/bg} = \sum^{N_{\rm dt/bg}}_{k}\ln\left[\frac{|\mathcal{M}(\xi^k)|^2}{\int\epsilon(\xi)|\mathcal{M}(\xi)|^2 R_{3}(\xi)d\xi}\right],
\end{equation}
where $\epsilon$ denotes the detection efficiency, $R_3$ is the three-body phase space factor, and $\xi^{k}$ is the measurement of the $k$-th event, namely the four-momenta of the final state $\gamma K_S^0 K_S^0$. The background contribution is modeled with the inclusive MC simulation and directly subtracted from the data sample. The integral in the denominator is numerically calculated with a MC method as
\begin{equation}
\int\epsilon(\xi)|\mathcal{M}(\xi)|^2 R_{3}(\xi)d\xi \propto \sum^{N_{\rm MC}}_{k_{\rm MC}}|\mathcal{M}(\xi^{k_{\rm MC}})|^2.
\end{equation}
Here, $N_{\rm MC}$ is the number of simulated
events uniformly distributed in the $\psi(3686)\to\gamma K_S^0 K_S^0$ phase space after detector simulation and event selection, which is 25 times larger than the data sample.

Technically, the calculation of the log-likelihood is accelerated by the GPU. The minimization of $-\ln\mathcal{L}$ is executed with the {\sc minuit} package~\cite{James:1975dr}.

\subsection{Significance test}

To determine which resonance states should be included in the nominal solution, several models $M_j$ with different sets of resonances are tried based on the moment distributions $\langle Y_l^0 \rangle$. Considering only the $f_0$ and $f_2$ contributions, the moments are related to the spin-0~(S) and spin-2~(D) amplitudes by~\cite{BESIII:2022iwi}
\begin{equation}
\begin{split}
\sqrt{4\pi} \langle Y_0^0 \rangle &= S_0^2+D_0^2+D_1^2+D_2^2, \\
\sqrt{4\pi} \langle Y_2^0 \rangle &= \frac{1}{7\sqrt{5}}(10 D_0^2+5 D_1^2-10 D^2_2) + 2 S_0 D_0\cos\phi_{D_0}, \\
\sqrt{4\pi} \langle Y_4^0 \rangle &= \frac{1}{7}(6D_0^2-4D_1^2+D_2^2), \\
\end{split}
\end{equation}
where $\phi_{D_0}$ is the phase of the $D$-wave relative to the $S$-wave. The moments without acceptance correction $ \langle Y_l^{\prime 0} \rangle= \sum^{N_{\rm dt}}_{i}{P_l(\cos\theta^{i}_{K_S^0 })}$ are extracted from data sample by re-weighting the mass spectrum $M_{K_S^0  K_S^0 }$ with the $l$-th Legendre polynomials $P_l$ after subtracting the simulated background sample, where $\theta_{K_S^0 }$ is the opening angle between the opposite flight directions of $\psi(3686)$ and $K_S^0$ in the rest frame of the $K^0_S K^0_S$ system. The determination of the nominal solution is described below and summarized in Table~\ref{tab:significance_test}.

\begin{figure*}[htbp]
	\centering
	\begin{minipage}[t]{0.32\textwidth}
		\centering
		\includegraphics[width=5.2cm]{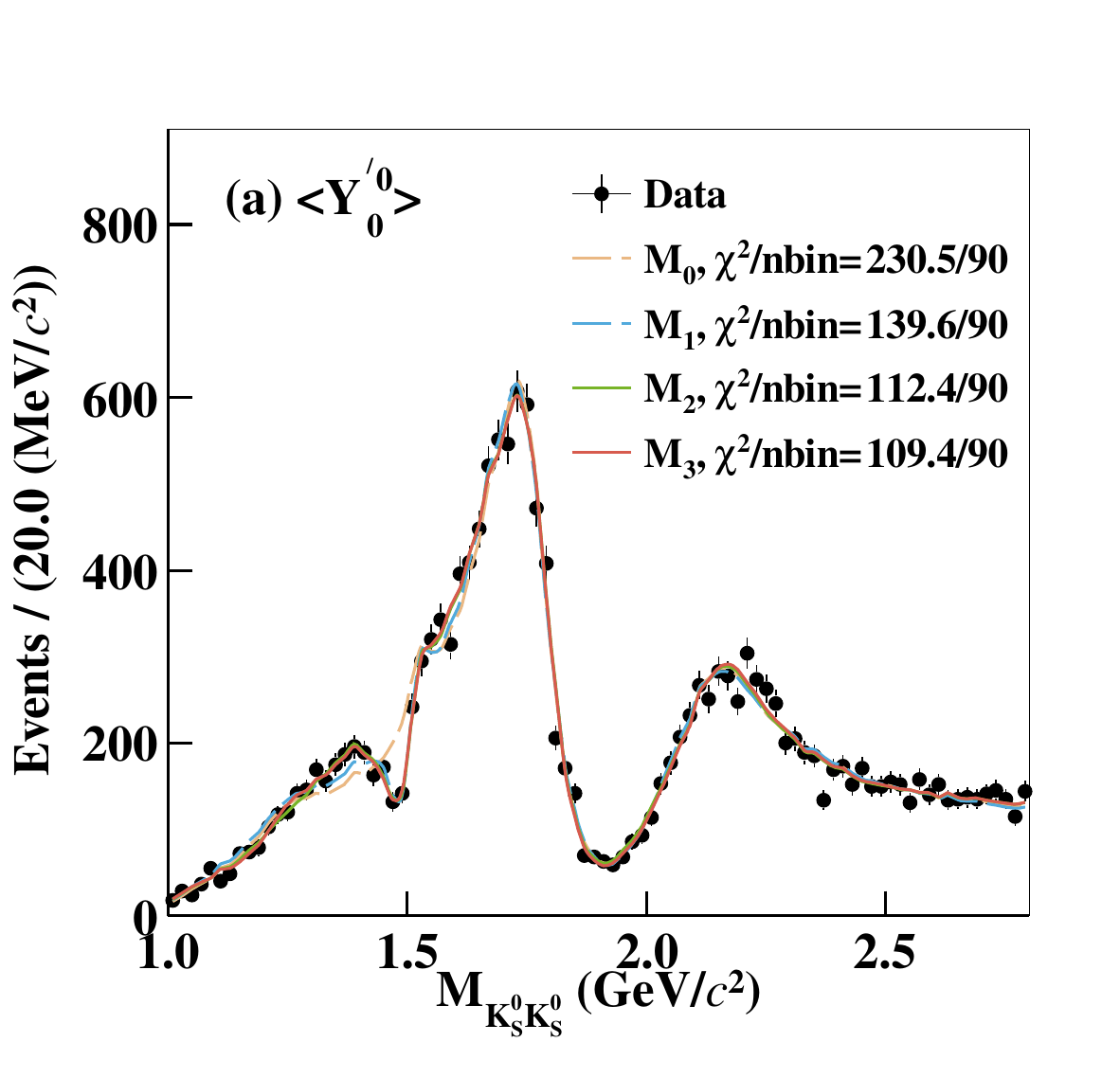}
	\end{minipage}
	\begin{minipage}[t]{0.32\textwidth}
		\centering
		\includegraphics[width=5.2cm]{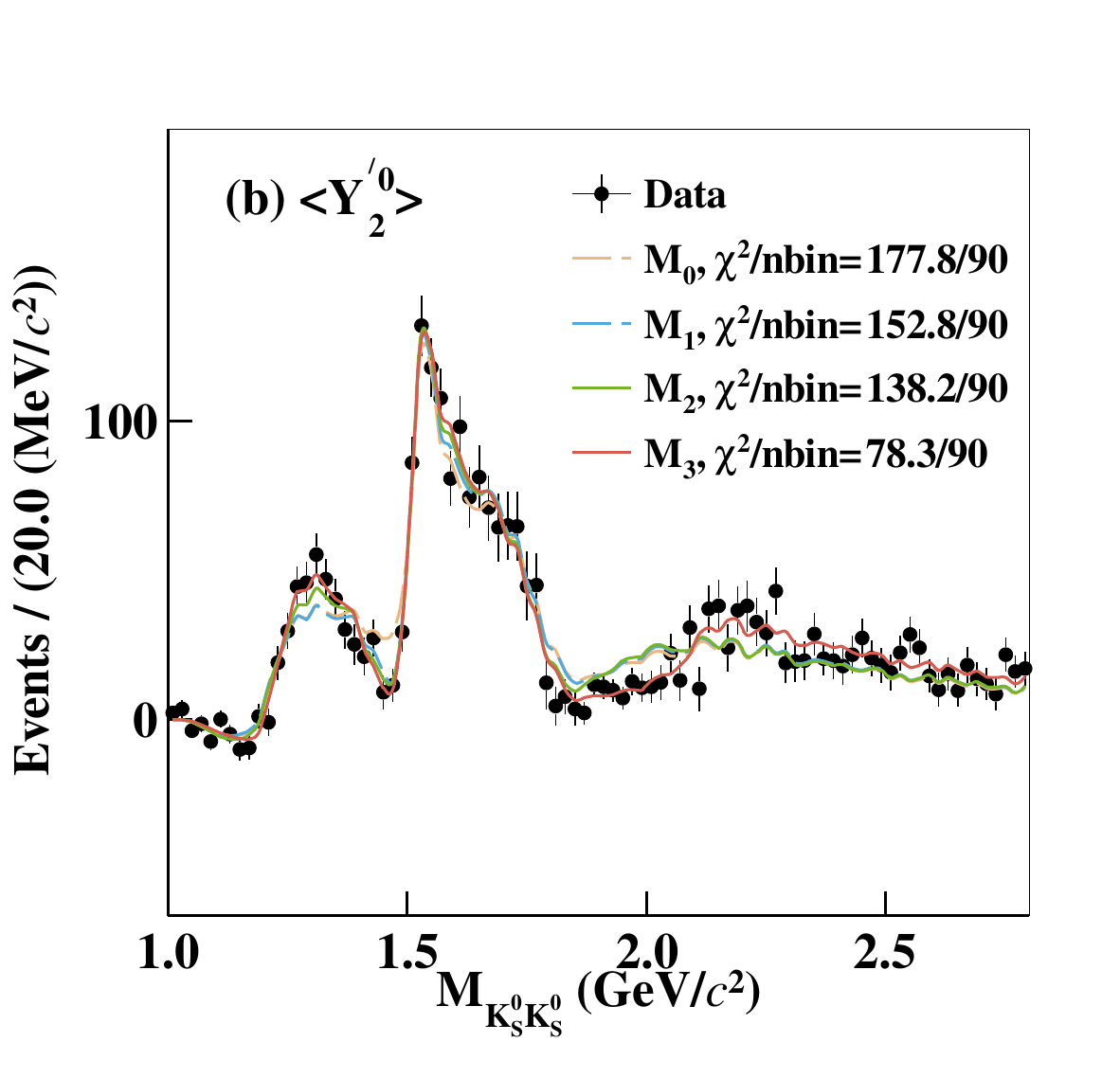}
	\end{minipage}
	\begin{minipage}[t]{0.32\textwidth}
		\centering
		\includegraphics[width=5.2cm]{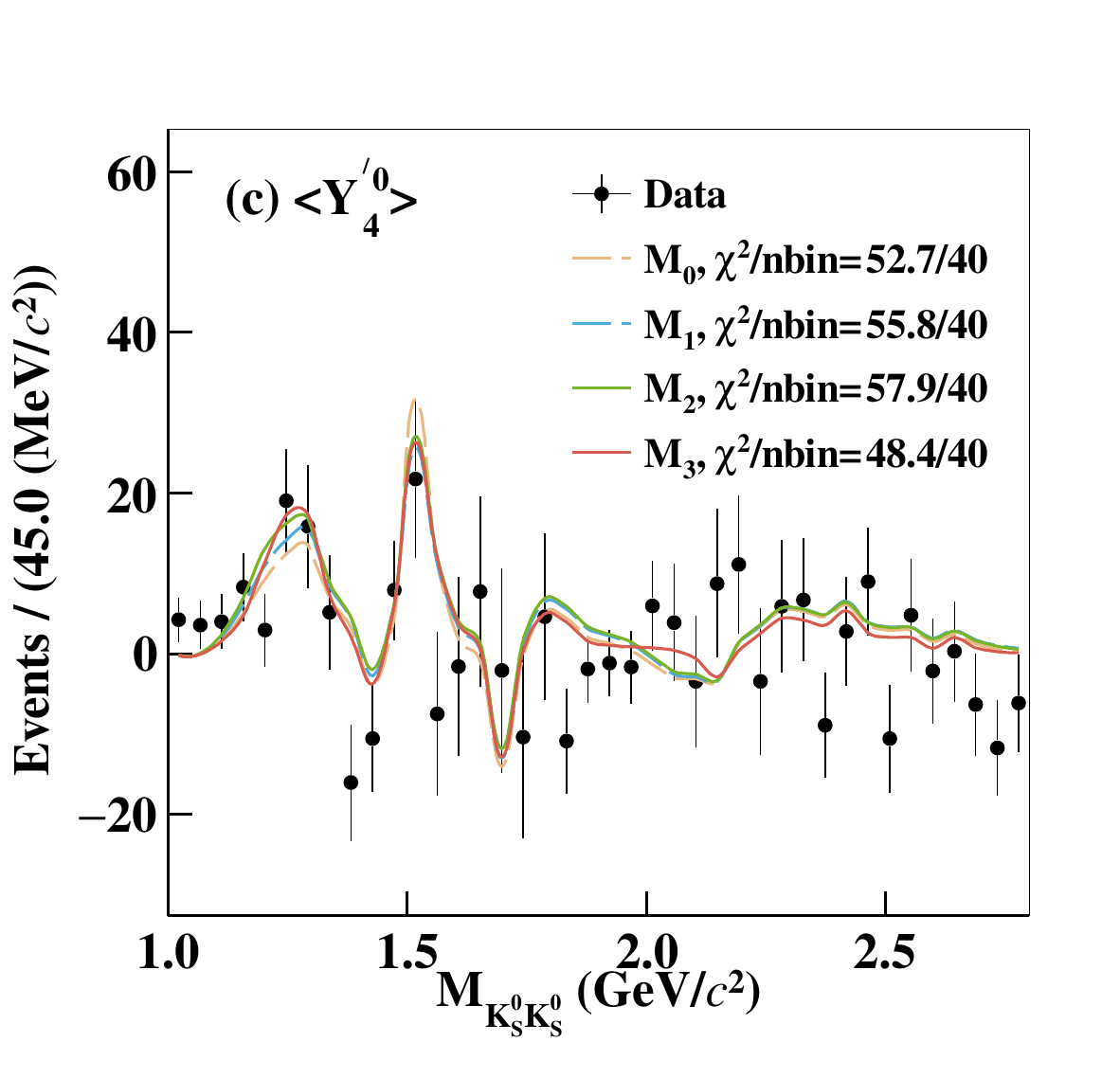}
	\end{minipage}  \\
	\caption{The moments of (a) $\langle Y_0^{\prime 0} \rangle$, (b) $\langle Y_2^{\prime 0} \rangle$, and (c) $\langle Y_4^{\prime 0} \rangle$ without acceptance correction. The black dots with error bars are the data sample after subtracting the simulated background sample. The curves in various colors are fit projections of different models.}
	\label{fig:angular_moments}
\end{figure*}

\begin{table}[htbp]
	\centering
	\tabcolsep=0.7cm
	\caption{The numbers of float parameters $N_{\rm par}$, log-likelihood values relative to the minimal model $\ln \mathcal{L}$, and the statistical significance of added poles. The poles included in the minimal model $M_0$ are $f_0(1710)$, $f_0(2020)$, $f_2(1270)$, and $f^\prime_2(1525)$.}
	\begin{tabular}{ccccc}
		\hline\hline
		Model & Add Pole &  $N_{\rm par}$ & $\ln \mathcal{L}$ & Significance\\
		\hline
		$M_0$ & --- & 26 & 0 & --- \\
		$M_1$ & $f_0(1500)$ & 30 & 56.8 & 10.0 \\
		$M_2$ & $f_0(1370)$ & 34 & 77.8 & 5.6 \\
		$M_3$ & $f_2(1950)$ & 42 & 133.9 & 9.3 \\
		\hline
		Scan $f_0$ & $f_0(X)$ & 46 & 137.5 &  $<3$\\
		Scan $f_2$ & $f_2(X)$ & 50 & 139.4 &  $<3$\\
		\hline
		\hline
	\end{tabular}
	\label{tab:significance_test}
\end{table}

\begin{itemize}
	\item As shown in Figure~\ref{fig:angular_moments}, several significant peaks in the $\langle Y_l^{\prime 0} \rangle(M_{K_S^0  K_S^0 })$ distributions are identified as $f_0(1710)$, $f_0(2020)$, $f_2(1270)$, and $f^\prime_2(1525)$. Thus, the minimal model $M_0$ is chosen to include these four states. It should be noted that one can not control which resonances are included directly. However, by imposing start-parameter ranges and parameters limits on the bare masses and couplings, the fitted pole positions do match the expected resonance well. Additionally, the $f_0$ background is added in the $\mathcal{P}$-vector to describe the non-resonant structures, while the $f_0$ background term in the K-matrix is found to be insignificant. The $K^*(892)^0\to K_S^0 \gamma$ contribution is included, as suggested by the $M_{K_S^0 \gamma}$ spectra in Figure~\ref{fig:fit_projection}. No evidence of $f_0(980)$ is observed near the $K\bar{K}$ threshold. The fit projection of $M_0$ is displayed in Figure~\ref{fig:angular_moments}, revealing significant deviations around 1.5 GeV/$c^2$ in the $\langle Y_{0(2)}^{\prime 0} \rangle$ distributions.
	
	\item Next, the $f_0(1500)$ is added into the model as $M_1$. This significantly improves the description of the dip around 1.5 GeV/$c^2$. With the change in the log-likelihood value $\Delta(\ln \mathcal{L})$ and the number of free parameters $\Delta N_{\rm par}$ relative to the previous model, the statistical significance of $f_0(1500)$ is calculated to be 10$\sigma$ using the Wilk's theorem as an approximate estimation. However, some deviation around 1.3 GeV/$c^2$ in the $\langle Y_{0(2)}^{\prime 0} \rangle$ spectrum remains.
	
	\item To address this, $f_0(1370)$ is included in the model as $M_{2}$, which provides a good description of the peak around $1.3$ GeV/$c^2$. The statistical significance of $f_0(1370)$ is calculated to be 5$\sigma$.
	
	\item The discrepancy around 2.0 GeV/$c^2$ on the $\langle Y_{2}^{\prime 0} \rangle$ spectrum is attributed to the absence of a broad $f_2$ contribution. There are two potential sources: an additional $f_2$ pole or $f_2$ background. It is found that including an additional $f_2$ pole provides a significantly better fit than the $f_2$ background with $\Delta \ln \mathcal{L}=40$. After adding one more $f_2$ pole around 2.0 GeV/$c^2$ as model $M_{3}$, all three moments $\langle Y_{L}^{\prime 0} \rangle$ are well described. The determined pole position is consistent with $f_2(1950)$ in the  PDG~\cite{ParticleDataGroup:2024cfk}. The statistical significance of this $f_2(1950)$ pole is determined to be 9$\sigma$.
	
	\item Finally, attempts are made to include an additional $f_{0,2}$ pole or $f_4$ resonance state into model $M_3$ by scanning its bare mass from 1.0 GeV to 2.5 GeV. Other parameters are kept to be float in this scan. None of them yields a statistical significance greater than 3$\sigma$. Therefore, the model $M_{3}$ is taken as the nominal solution.
\end{itemize}

\subsection{Nominal solution}

\begin{figure*}[htbp]
	\centering
	\centering
	\includegraphics[width=15.0cm]{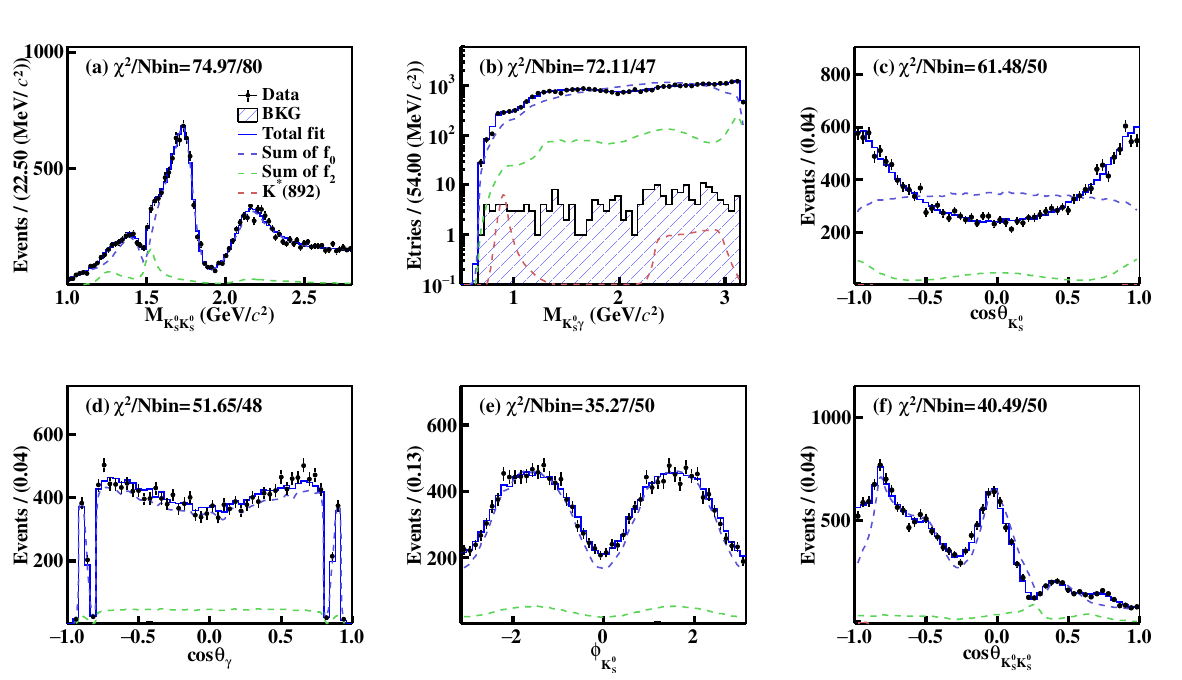}
	\caption{The fit projections of the nominal solution. The black dots with error bars are the data sample. The solid blue curve is the fit result. The hatched histogram is the background. The dashed lines with various colors are different components. The projections from the two identical $K^0_S$ mesons are merged in subplot~(b).}
	\label{fig:fit_projection}
\end{figure*}

Figure~\ref{fig:fit_projection} shows the fit projections of nominal solution on the two-body invariant mass spectra and several helicity angular distributions. In addition to the previously defined $\theta_{K_S^0}$, $\theta_\gamma$ represents the flight direction of $\gamma$ in the rest frame of $\psi(3686)$; $\phi_{K^0_S}$ is the angle between the decay planes of $\psi(3686)\to\gamma f$ and $f\to K_S^0 K_S^0$ in the rest frame of $\psi(3686)$; and $\theta_{K^0_S K^0_S}$ is the opening angle between the two $K_S^0$ flight directions in the rest frame of $\psi(3686)$. The nominal solution provides a good description of the data sample.

\subsection{Pole positions}

The pole positions are obtained by solving the complex function $(1-\mathcal{K}i\rho n^2)=0$ numerically in the unphysical sheet ${\rm Im}(k)<0$, where the $k$ is the momentum of $K_S^0 $ in the rest frame of $f_{0,2}$. The statistical uncertainty is accessed by generating 1000 sets of K-matrix parameters based on the corresponding covariance matrix and calculating the pole positions. The standard deviations of the calculated pole positions are taken to be statistical uncertainties. The results are summarized in Table~\ref{tab:poles}. The reliability of statistical uncertainties are further validated with the bootstrap method~\cite{Langenbruch:2019nwe} by performing fits to alternative data samples obtained by sampling with replacement from original data sample by 100 times. All of the alternative samples share the same statistics as the original one. The standard deviations of the resultant distributions of pole positions obtained from fits to alternative data samples, are found to be consistent with the statistical uncertainties obtained in the nominal approach.

\begin{table}[htbp]
	\centering
	\renewcommand{\arraystretch}{1.2}
	\caption{The solved pole positions in MeV and corresponding squared modulus of residues $|\mathcal{R}|^2$ in $(\times 10^{-4})$. Here, the uncertainty is statistical only.}
	\begin{tabular}{cccccc}
		\hline\hline
		Resonance & Pole positions & $|\mathcal{R}_{E_1}|^2$ & $|\mathcal{R}_{M_2}|^2$ & $|\mathcal{R}_{E_3}|^2$ &  $|\mathcal{R}_{\rm sum}|^2$\\
		\hline
		$f_0(1370)$ & ($1297.4\pm56.9$)$-i(125.8\pm71.5$) & $19.1^{+45.6}_{-12.1}$ & ---  & --- & $19.1^{+45.6}_{-12.1}$
		\\
		$f_0(1500)$ & ($1487.8\pm7.7$)$-i(46.1\pm8.6$) & $11.9^{+9.7}_{-4.9}$ & ---  & --- & $11.9^{+9.7}_{-4.9}$
		\\
		$f_0(1710)$ & ($1770.7\pm5.1$)$-i(84.7\pm3.8$) & $274.2^{+40.2}_{-29.9}$ & ---  & --- & $274.2^{+40.2}_{-29.9}$
		\\
		$f_0(2100)$ & ($2141.9\pm17.2$)$-i(138.8\pm10$) & $137.3^{+24.5}_{-22.0}$ & ---  & --- & $137.3^{+24.5}_{-22.0}$
		\\
		$f_2(1270)$ & ($1226.2\pm8.6$)$-i(78.9\pm7.5$) & $7.4^{+2.0}_{-2.4}$ & $9.3^{+2.2}_{-2.7}$ & $3.3^{+1.9}_{-1.0}$ & $20.1^{+3.9}_{-4.4}$\\
		$f_2(1525)$ & ($1515.7\pm4.3$)$-i(33.7\pm3.8$) & $3.4^{+1.8}_{-1.6}$ & $3.2^{+1.1}_{-1.2}$ & $0.7^{+1.0}_{-0.4}$ & $7.3^{+2.6}_{-2.1}$\\
		$f_2(1950)$ & ($2069.9\pm22.2$)$-i(130\pm26$) & $25.5^{+17.1}_{-11.5}$ & $5.5^{+5.5}_{-3.1}$ & $2.6^{+3.9}_{-1.5}$ & $33.5^{+20.7}_{-12.7}$\\
		\hline
		\hline
	\end{tabular}
	\label{tab:poles}
\end{table}

\subsection{Pole residues}

The contributions of individual $f_{0,2}$ poles $a$ in the decay $\psi(3686)\to\gamma K_S^0 K_S^0$ are quantified with the residue of $f^{(f_{0,2})}_i=n(1-\mathcal{K}i\rho n^2)^{-1}\mathcal{P}_{i}$ in the complex-$s$ plane at their pole positions $s_a$. The residues are proportional to the product of the production and decay strength, hence the combined branching fraction of $\psi(3686)\to\gamma f_{0,2},f_{0,2}\to K_S^0 K_S^0$ is given as
\begin{equation}
\mathcal{B}\left(\psi(3686)\to\gamma f_{0,2},f_{0,2}\to K_S^0 K_S^0\right) \propto |\mathcal{R}|^2,
\end{equation}
where the residue $\mathcal{R}$ is defined as
\begin{equation}
\begin{split}
\mathcal{R}(s_a) =& \frac{1}{2\pi i}\oint \sqrt{\rho_{\psi\to\gamma f}(s_a)}\times f^{(f_{0,2})}_i(s_a)\times \sqrt{\rho_{f\to K_S^0 K_S^0}(s_a)} ds \\
=& \frac{1}{2\pi i}\int_{0}^{2\pi} \sqrt{\rho_{\psi\to\gamma f}(s_a+re^{i\theta})}\times f^{(f_{0,2})}_i(s_a+re^{i\theta})\times \sqrt{\rho_{f\to K_S^0 K_S^0}(s_a+re^{i\theta})} ire^{i\theta}d\theta
\end{split}
\label{eq:res_def}
\end{equation}
Here, we use $ds=d(s_a+re^{i\theta})=ire^{i\theta}d\theta$ and $r$ should be small enough to ensure only one pole $s_a$ in this circle contour $s_a+re^{i\theta}$. This integral is calculated numerically via
\begin{eqnarray}
\mathcal{R}(s_a) \simeq \frac{1}{2\pi i}\sum_{n=1}^{N} \sqrt{\rho_{\psi\to\gamma f}(s_n)}\times f^{(f_{0,2})}_i(s_n)\times \sqrt{\rho_{f\to K_S^0 K_S^0}(s_n)} \times ir e^{in\Delta\theta} \times \Delta\theta,
\end{eqnarray}
where $\Delta \theta=2\pi/N$ and $s_n = s_a+re^{in\Delta\theta}$. However, it should be noted the covariant tensor amplitudes used in the amplitude fit are not well normalized and orthogonal. Therefore, the obtained residues $\mathcal{R}^{f_{0,2}}_{\rm cov}$ in covariant tensor amplitude are further converted to the residues $\mathcal{R}^{f_{0,2}}_{\rm E1/M2/E3}$ in helicity amplitude in multi-pole basis, which is written as~\cite{BESIII:2015rug}
\begin{equation}
\begin{split}
I &= \sum_{\lambda_\psi=\lambda_{\gamma} = \pm 1} |h_0(\theta_{K_S^0})d^1_{1,1}(\theta_\gamma)e^{\pm i\phi_{K_S^0}}+h_1(\theta_{K_S^0})d^1_{1,0}(\theta_\gamma)+h_2(\theta_{K_S^0})d^1_{1,-1}(\theta_\gamma)e^{\mp i \phi_{K_S^0}}|^2 \\
&+\sum_{\lambda_\psi=-\lambda_{\gamma} = \pm 1} |h_0(\theta_{K_S^0})d^1_{1,-1}(\theta_\gamma)e^{\mp i\phi_{K_S^0}}-h_1(\theta_{K_S^0})d^1_{1,0}(\theta_\gamma)+h_2(\theta_{K_S^0})d^1_{1,1}(\theta_\gamma)e^{\pm i \phi_{K_S^0}}|^2,
\end{split}
\end{equation}
where the $h_{0,1,2}$ is defined as
\begin{equation}
\begin{split}
h_0(\theta_{K_S^0}) =& \sqrt{3}\alpha_{01}+\sqrt{\frac{3}{2}}(\alpha_{21}+\sqrt{5}\alpha_{22}+2\alpha_{23}) d^2_{0,0} (\theta_{K_S^0}) \\
h_1(\theta_{K_S^0}) =&\frac{\sqrt{2}}{2}(3\alpha_{21}+\sqrt{5}\alpha_{22}-4\alpha_{23}) d^2_{1,0} (\theta_{K_S^0}) \\
h_2(\theta_{K_S^0}) =& (3\alpha_{21} -\sqrt{5}\alpha_{22} + \alpha_{23}) d^2_{2,0} (\theta_{K_S^0}).
\end{split}
\end{equation}
Here, $\alpha_{01}$ is the complex coefficient of E1 transition of $0^{++}$; $\alpha_{21}$, $\alpha_{22}$, and $\alpha_{23}$ are the complex coefficients of E1, M2, and E3 transitions of $2^{++}$, respectively. Following Refs.~\cite{BESIII:2015rug,zou:2004Angular}, the conversion relationship for $\psi(3686)\to\gamma f_0$ reads $\mathcal{R}^{f_0}_{\rm E_1} = \sqrt{\frac{1}{6}} \mathcal{R}^{f_0}_{\rm cov}$ and the relationship for $\psi(3686)\to\gamma f_2$ reads
\begin{equation}
\begin{bmatrix}
\mathcal{R}^{f_2}_{\rm E_1} \\
\mathcal{R}^{f_2}_{\rm M_2} \\
\mathcal{R}^{f_2}_{\rm E_3}
\end{bmatrix}
=
\begin{bmatrix}
\sqrt{\frac{3}{2}} & \sqrt{\frac{15}{2}} & 2\sqrt{\frac{3}{2}} \\
3\sqrt{\frac{1}{2}} & \sqrt{\frac{5}{2}} & -2\sqrt{2} \\
3 & -\sqrt{5} & 1
\end{bmatrix}^{-1}
\times
\frac{2\sqrt{2}}{3}Q^2_{K^0_S}
\times
\begin{bmatrix}
-1 & - 2 \frac{E_\gamma^2}{m^2_{f_2}} M_\psi^2 & 0 \\
-\sqrt{3}\frac{E_{f_2}}{m_{f_2}} & 0 & \sqrt{3}\frac{E_\gamma^2}{m_{f_2}}  M_\psi \\
-\sqrt{6} & 0 & 0
\end{bmatrix}
\times
\begin{bmatrix}
\mathcal{R}^{f_2}_{\rm cov,1} \\
\mathcal{R}^{f_2}_{\rm cov,2} \\
\mathcal{R}^{f_2}_{\rm cov,3}
\end{bmatrix},
\label{eq:f2_conversion}
\end{equation}
where the $E_{\gamma,f_2}$ are defined in the $\psi(3686)$ rest frame and $Q_{K^0_S}$ is the momentum of $K_S^0$ in the $K^0_S K^0_S$ rest frame. All the kinematic variables shown in Eq.~\ref{eq:f2_conversion} are calculated at the pole masses of the $f_2$ states $m_{K^0_S K^0_S}=m_{f_2}$. The total contribution of $f_2$ state $|\mathcal{R}_{\rm sum}|^2$ is calculated with $|\mathcal{R}_{E_1}|^2+|\mathcal{R}_{M_2}|^2+|\mathcal{R}_{E_3}|^2$. The obtained results are also summarized in Table~\ref{tab:poles}.

\section{Systematic uncertainties}

The systematic uncertainties of the amplitude analysis are summarized in Table~\ref{tab:sum_sys_amp}. They are estimated as follows. The total systematic uncertainty is obtained by adding the individual ones in quadrature.

\begin{table*}[htbp]
	\centering
	\caption{The systematic uncertainties for pole positions in MeV and pole residue $|\mathcal{R}|^2$ in $(\times 10^{-4})$.}
	\scalebox{0.68}{
		\begin{tabular}{c|c|cccccccc}
			\hline
			\hline 
			\multicolumn{2}{c|}{Pole properties} & Hadron scale  & Fit bias & $M_{K_S^0  K_S^0 }$ calibration & $M_{K_S^0  K_S^0 }$ cut & Helix corr & Add pole & $\mathcal{P}$-vector BG &  Sum \\ \hline
			\multirow{7}{*}{Real} & $f_0(1370)$ & 4.3 & 20.4 & \multirow{7}{*}{2.2} & 11.3 & 3.1 & 8.3 & 14.8 & 29.4 \\
			& $f_0(1500)$ & 0.5 & 0.2 &   & 1.5 & 0.6 & 2.3 & 0.7 & 3.7 \\
			& $f_0(1710)$ & 0.1 & 0.0 &   & 2.2 & 0.7 & 2.1 & 1.7 & 4.2 \\
			& $f_0(2020)$ & 1.5 & 3.0 &   & 9.8 & 3.9 & 11.1 & 8.3 & 17.9 \\
			& $f_2(1270)$ & 6.9 & 1.4 &   & 3.0 & 0.2 & 0.3 & 3.8 & 8.8 \\
			& $f_2(1525)$ & 1.0 & 2.1 &   & 0.3 & 0.2 & 0.7 & 0.2 & 3.3 \\
			& $f_2(1950)$ & 3.7 & 13.4 &   & 5.6 & 2.1 & 3.0 & 8.3 & 17.6 \\
			\hline
			\multirow{7}{*}{Imag} & $f_0(1370)$ & 2.1 & 2.0 & 0.1 & 19.5 & 5.3 & 15.3 & 35.9 & 44.0 \\
			& $f_0(1500)$ & 0.7 & 4.2 & 0.1 & 1.9 & 0.5 & 3.5 & 3.2 & 6.6 \\
			& $f_0(1710)$ & 0.6 & 0.3 & 0.1 & 0.2 & 0.2 & 4.0 & 2.4 & 4.7 \\
			& $f_0(2020)$ & 0.7 & 3.1 & 0.1 & 4.6 & 0.3 & 0.6 & 13.3 & 14.4 \\
			& $f_2(1270)$ & 11.1 & 1.5 & 0.1 & 2.4 & 0.8 & 2.3 & 0.4 & 11.7 \\
			& $f_2(1525)$ & 0.4 & 3.3 & 0.2 & 0.6 & 1.4 & 1.6 & 1.0 & 4.1 \\
			& $f_2(1950)$ & 18.7 & 0.7 & 0.1 & 2.2 & 0.7 & 4.0 & 1.5 & 19.3 \\
			\hline
			\multirow{16}{*}{$|\mathcal{R}|^{2}$} & $f_0(1370)$ & 0.7 & 0.4 & \multirow{16}{*}{---} & 17.1 & 1.6 & 14.4 & 14.4 & 26.7 \\
			& $f_0(1500)$ & 0.5 & 3.6 &   & 2.8 & 0.1 & 3.2 & 3.8 & 6.8 \\
			& $f_0(1710)$ & 1.1 & 5.9 &   & 17.3 & 3.2 & 19.6 & 4.7 & 27.4 \\
			& $f_0(2020)$ & 0.6 & 2.7 &   & 13.6 & 0.2 & 10.1 & 33.0 & 37.1 \\
			& $f_2(1270)$, $E_1$ & 3.1 & 0.0 &  & 1.1 & 0.2 & 0.7 & 1.3 & 3.6 \\
			& $f_2(1270)$, $M_2$ & 2.2 & 0.8 &  & 0.8 & 0.3 & 0.5 & 1.3 & 2.9 \\
			& $f_2(1270)$, $E_3$ & 0.8 & 0.3 &  & 0.2 & 0.0 & 1.0 & 0.4 & 1.4 \\
			& $f_2(1270)$, Sum & 6.1 & 0.4 &  & 1.8 & 0.5 & 2.2 & 3.0 & 7.4 \\
			& $f_2(1525)$, $E_1$ & 0.7 & 0.0 &  & 0.2 & 0.2 & 0.7 & 0.0 & 1.1 \\
			& $f_2(1525)$, $M_2$ & 0.1 & 0.4 &  & 0.1 & 0.2 & 0.8 & 0.3 & 1.0 \\
			& $f_2(1525)$, $E_3$ & 0.1 & 0.1 &  & 0.1 & 0.2 & 0.3 & 0.1 & 0.4 \\
			& $f_2(1525)$, Sum & 0.7 & 0.3 &  & 0.3 & 0.5 & 1.8 & 0.4 & 2.1 \\
			& $f_2(1950)$, $E_1$ & 7.0 & 0.1 &  & 5.2 & 0.4 & 7.3 & 3.4 & 11.9 \\
			& $f_2(1950)$, $M_2$ & 2.0 & 2.0 &  & 0.1 & 0.3 & 2.0 & 0.8 & 3.6 \\
			& $f_2(1950)$, $E_3$ & 0.8 & 0.6 &  & 0.7 & 0.1 & 0.7 & 0.4 & 1.5 \\
			& $f_2(1950)$, Sum & 9.8 & 2.3 &  & 5.0 & 0.8 & 10.0 & 4.6 & 15.8 \\
			\hline
			\hline
		\end{tabular}
	}
	\label{tab:sum_sys_amp}
\end{table*}

The systematic uncertainty associated with the choice of the hadron ``scale'' parameter $Q_0$ is estimated by performing scan on $Q_0$ value. The difference between the nominal solution with $Q_0=3$ GeV$^{-1}$ and alternative fit with the best $Q_0=2.2$ GeV$^{-1}$ is taken as the systematic uncertainty.

The systematic uncertainty due to fit bias is estimated with a toy MC study. A total of 100 sets of signal MC samples are generated based on the nominal solution with the background contribution mixed. The same amplitude fits are performed on these pseudo-data samples, and any observed fit biases are assigned as the systematic uncertainty.

The systematic uncertainty of the mass calibration is studied by performing a 1D fit to the $M_{K_S^0 K_S^0 }$ distribution of the data sample around the $\chi_{c0}$ mass region. The resolution and mass shift of $M_{K_S^0  K_S^0 }$ are determined to be $\sigma=7.3$ MeV/$c^2$ and $\Delta M =2.2$ MeV/$c^2$, respectively. As a conservative estimation, the corresponding impacts on the pole positions $\Re-i\Im$ are assigned as $\Delta \Re =2.2$ MeV/$c^2$ and $\Delta \Im =(\sqrt{(\Im^2+(\sigma/2)^2)}-|\Im|)$.

The uncertainty caused by the cut $M_{K_S^0  K_S^0 }<2.8$ GeV/$c^2$ is estimated by varying the cut range to $M_{K_S^0  K_S^0 }<2.7$ GeV/$c^2$ and 2.9 GeV/$c^2$ and performing alternative fits. The maximum changes on pole positions are taken as the systematic uncertainties.

The systematic uncertainty associated with the 4C kinematic fit is studied by performing an alternative amplitude fit with a phase space MC sample after helix correction~\cite{BESIII:2012mpj}. The difference between this alternative fit and nominal fit is taken as the systematic uncertainty.

The systematic uncertainty due to additional $f_{0,2,4}$ state is assigned as the largest difference between the fit results of nominal solution and the solution with the largest improvement on $\ln\mathcal{L}$ in the scan of additional $f_{0,2}$ pole and $f_4$  resonance state.

The systematic uncertainty due to background modeling in $\mathcal{P}$ vector is estimated by using the first-order polynomial background term as $\mathcal{P}_{i} = \sum_{a} \frac{\beta^{i}_{a} g_a}{m_a^2-s}+\beta^{i,0}_{\rm bkg}+\beta^{i,1}_{\rm bkg}\times s$ in alternative fit. The difference in pole positions between nominal and alternative fit is taken as the systematic uncertainty.

\section{The branching fractions of individual poles}

The branching fractions of individual poles are calculated as
\begin{equation}
\mathcal{B}(\psi(3686)\to\gamma R, R\to K_S^0 K_S^0)=\sigma_{a}/\sigma_{\rm tot}\times\mathcal{B}^{\prime}(\psi(3686)\to\gamma K_S^0 K_S^0).
\end{equation}
Here, $\mathcal{B}^{\prime}(\psi(3686)\to\gamma K_S^0 K_S^0)$ is the total branching fraction with $M_{K_S^0 K_S^0}<2.8$~GeV/$c^2$; $\sigma_{a}$ and $\sigma_{\rm tot}$ are the total decay rates of pole $a$ and the process, respectively. The total decay rate $\sigma_{\rm tot}$ is calculated in a numerical approach as
\begin{equation}
\sigma_{\rm tot} = \int{|\mathcal{M}|^2 d\Phi_3} \simeq |\mathcal{\bar{M}}|^2 \times \Phi_3,
\end{equation}
where $|\mathcal{\bar{M}}|^2$ is the averaged amplitude squared of phase space MC sample without detector efficiency and $\Phi_3$ is the three-body phase space factor of $\psi(3686)\to\gamma K_S^0 K_S^0$ with $M_{K_S^0 K_S^0}<2.8$ GeV/$c^2$. The $\Phi_3$ is calculated with 
\begin{equation}
\Phi_3 = \int^{2.8^2}_{4M^2_{K^0_S}} \rho_{\psi(3686)\to\gamma a}(s)\times\rho_{a\to K_S^0 K_S^0}(s) ds,
\end{equation}
where $\rho$ indicates corresponding two-body phase space factor. For the decay rate of pole $a$, one can not intuitively separate the contribution of individual pole from the total amplitude. Therefore, as an approximate approach, a Breit-Wigner function $\mathcal{R}^{a}_{\rm sum}/(s-M_a^2+iM_a\Gamma_a)$, which yields exactly the same pole position $M_a-i\Gamma_a/2$ and residue $\mathcal{R}^{a}_{\rm sum}$, is used to model the contribution from individual pole $|\mathcal{M}_{a}|^2$. The decay rate of such a Breit-Wigner function is calculated to be 
\begin{equation}
\sigma_{a} = \int^{2.8^2}_{4M^2_{K_S^0}} |\frac{\mathcal{R}^{a}_{\rm sum}}{s-M_a^2+iM_a\Gamma_a}|^2 ds\simeq\int^{\infty}_{0} |\frac{\mathcal{R}^{a}_{\rm sum}}{s-M_a^2+iM_a\Gamma_a}|^2 ds=\frac{\pi |\mathcal{R}^{a}_{\rm sum}|^2}{M_a\Gamma_a}.
\end{equation}
Here, the narrow width assumptions~\cite{VonDetten:2021rax} are used and it should be noted that the phase space factor has been included in the residue following Eq.~\ref{eq:res_def}.

The total branching fraction $\mathcal{B}^{\prime}(\psi(3686)\to\gamma K_S^0 K_S^0)$ is measured to be
\begin{equation}
\mathcal{B}^{\prime}(\psi(3686)\to\gamma K_S^0 K_S^0) = \frac{N_{\rm sig}}{N_{\psi(3686)}\times\epsilon\times\mathcal{B}^2(K_S^0\to\pi^+\pi^-)}=(5.91\pm0.05)\times 10^{-5}
\end{equation}
where $N_{\rm sig} = 17567\pm134$ is the signal number obtained by counting; $N_{\psi(3686)}=(2712\pm15)\times10^{6}$ is the total $\psi(3686)$ number~\cite{BESIII:2024lks}; $\epsilon=(22.92\pm0.03)\%$ is the detection efficiency determined with signal MC sample generated based on the amplitude fit result; $\mathcal{B}(K_S^0\to\pi^+\pi^-)=(69.20\pm0.05)\%$ is quoted from PDG~\cite{ParticleDataGroup:2024cfk}. The systematic uncertainties of branching fraction $\mathcal{B}^{\prime}(\psi(3686)\to\gamma K_S^0 K_S^0)$ include the following sources
\begin{itemize}
	\item $\psi(3686)$ number: The systematic uncertainty of $\psi(3686)$ number is 0.6\%~\cite{BESIII:2024lks}.
	\item Photon reconstruction: The systematic uncertainty due to photon reconstruction efficiency is studied with the control sample $e^+e^-\to\gamma\mu^+\mu^-$, which is found to be 0.5\% per photon.
	\item $K_S^0$ reconstruction: The systematic uncertainty associated with the $K_S^0$ reconstruction efficiency is studied with the control samples $J/\psi\to K_S^0 K^+\pi^-$ and $J/\psi\to\phi K_S^0 K^+\pi^-$, which is found to be 0.5\% per $K_S^0$.
	\item MC statistic: The systematic uncertainty caused by MC statistic $\sqrt{\frac{1-\epsilon}{N\cdot \epsilon}}=0.1\%$.
	\item Helix correction: The systematic uncertainty associated with 4C kinematic fit is studied by performing helix correction to the signal MC sample. The variation on the signal efficiency 2.2\% is assigned as this uncertainty.
	\item The quoted $\mathcal{B}(K_S^0\to\pi^+\pi^-)$: The systematic uncertainty due to quoted $\mathcal{B}(K_S^0\to\pi^+\pi^-)=(69.20\pm0.05)\%$ is $0.1\%$.
\end{itemize}
The total systematic uncertainty are obtained to be 2.5\% by summing all sources quadratically. 

The branching fraction $\mathcal{B}^{\prime}(\psi(3686)\to\gamma K_S^0 K_S^0)$ is finally calculated as $(5.91\pm0.05\pm0.02)\times 10^{-5}$. The branching fractions of individual poles are summarized in Table~\ref{tab:poles_res_sum}.

\section{Summary}

By analyzing $(2712\pm 14)\times 10^6$ $\psi(3686)$ events, we have performed the first amplitude analysis of $\psi(3686)\to\gamma K_S^0  K_S^0 $. Based on a one-channel K-matrix approach for the dynamics description, the data sample is well described with four poles for the $f_0$-wave and three poles for the $f_2$-wave. 

\begin{figure*}[htbp]
	\centering
	\centering
	\includegraphics[width=16cm]{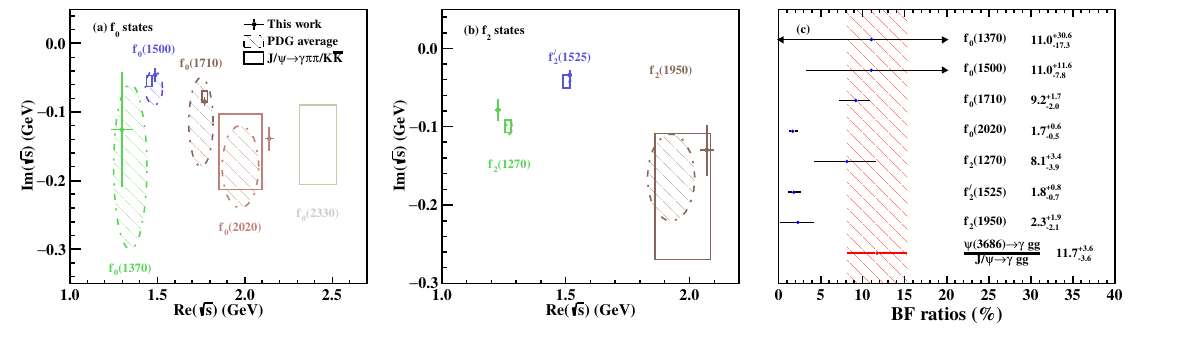}
	\caption{The pole positions of (a) $f_0$ and (b) $f_2$ states measured in this work, given by the PDG average, and $J/\psi\to\gamma\pi\pi/K\bar{K}$~\cite{Rodas:2021tyb}. The dots with error bars are obtained in this work. The hatched ellipses with dashed lines are the PDG values. The boxes with solid lines are cited from the 3-channel results of $J/\psi\to\gamma\pi\pi/K\bar{K}$~\cite{Rodas:2021tyb}, where $f_0(2330)$ could only be identified with the $4\pi$ channel included. (c) The branching fraction ratios ${\mathcal{B}(\psi(3686)\to\gamma f_{0,2})}/{\mathcal{B}(J/\psi\to\gamma f_{0,2})}$ and their comparison with ${\mathcal{B}(\psi(3686)\to\gamma gg)}/{\mathcal{B}(J/\psi\to\gamma gg)}$~\cite{ParticleDataGroup:2024cfk}.}
	\label{fig:pole_comparison}
\end{figure*}

The determined pole positions are summarized in Table~\ref{tab:poles_sum}, all of which match the well-known resonance states listed in the PDG as shown in Figure~\ref{fig:pole_comparison}. A further comparison is made with the combined analysis of $J/\psi\to\gamma\pi^0\pi^0/K^0_SK^0_S$~\cite{Rodas:2021tyb} as displayed in Figure~\ref{fig:pole_comparison}. The observations of $f_0(1500)$, $f_0(1710)$, $f^\prime_2(1525)$, and $f_2(1950)$ are in good agreement between the two studies. For the $f_0(1370)$, the determined pole position in this work is not stable, and no pole is identified as $f_0(1370)$ in Ref.~\cite{Rodas:2021tyb}. The observed $f_0$ pole above 2 GeV is assigned as $f_0(2020)$, the mass of which is slightly higher than the corresponding averaged value in PDG. This possibly be due to the fact that PDG assigns the pole with higher mass as $f_0(2100/2200)$, which may be the same state as $f_0(2020)$. The pole $f_0(2330)$ is not observed in this work, nor in the 2-channel fit in Ref.~\cite{Rodas:2021tyb}, but it is observed when including an additional $4\pi$ channel~\cite{Rodas:2021tyb}. The pole position of $f_2(1270)$ slightly deviates from the previous measurements potentially due to the absence of $\pi\pi$ couple-channel effect in this analysis. This comparison suggests that the produced $f_{0,2}$ poles in the $\psi(3686)$ radiative decay are in agreement with these in $J/\psi$ radiative decay.

The production behaviors of $f_0$ and $f_2$ poles in $\psi(3686)\to\gamma K_S^0 K_S^0$ are qualified with their residues in the multipole basis of helicity amplitude. The combined branching fractions $\mathcal{B}(\psi(3686)\to\gamma f_{0,2},f_{0,2}\to K_S^0 K_S^0)$ are also calculated with residues under narrow width assumption. No significant violation of the relationship $E_1>M_2>E_3$, which is predicted by the conventional quark model~\cite{Rodas:2021tyb} and confirmed in $J/\psi\to\gamma f_2$, is seen for $\psi(3686)\to\gamma f_2$ in current statistics. The branching fraction ratios ${\mathcal{B}(\psi(3686)\to\gamma f_{0,2})}/{\mathcal{B}(J/\psi\to\gamma f_{0,2})}$ are also listed in Table~\ref{tab:poles_res_sum} and shown in Figure~\ref{fig:pole_comparison}. By comparing the ratios with $\frac{\mathcal{B}(\psi(3686)\to\gamma gg)}{\mathcal{B}(J/\psi\to\gamma gg)}=(11.7\pm3.6)\%$, this work provides crucial experimental inputs on the internal structure of the $f_{0,2}$ states, especially their potential mixing with glueball components.

\begin{table*}[htbp]
	\centering
	\tabcolsep=0.15cm
	\caption{The pole positions determined in this work, cited from the PDG, and extracted from the 3-channel results of $J/\psi\to\gamma\pi\pi/K\bar{K}$~\cite{Rodas:2021tyb}. For this work, the first uncertainties are statistical and the second systematic. It should be noted that the PDG averaged values of some resonances have included Ref.~\cite{Rodas:2021tyb} hence there are some correlations.}
	\scalebox{0.75}{
		\begin{tabular}{cccc}
			\hline\hline 
			Resonance & This work & PDG & $J/\psi\to\gamma\pi\pi/K\bar{K}$~\cite{Rodas:2021tyb}\\
			\hline
			$f_0(1370)$ & $(1297.4\pm56.9\pm29.4)-i(125.8\pm71.5\pm44.0)$ & $(1250-1440)-i(60-300)$ &  ---\\
			$f_0(1500)$ & $(1487.8\pm7.7\pm3.7)-i(46.1\pm8.6\pm6.6)$ & $(1430-1530)-i(40-90)$ & $(1437-1471)-i(46-63)$\\
			$f_0(1710)$ & $(1770.7\pm5.1\pm4.2)-i(84.7\pm3.8\pm4.7)$ & $(1680-1820)-i(50-180)$ & $(1756-1785)-i(69-85)$\\
			$f_0(2020)$ & $(2141.9\pm17.2\pm17.9)-i(138.8\pm10.0\pm14.4)$ & $(1870-2080)-i(120-240)$ & $(1955-2098)-i(103-213)$ \\
			$f_2(1270)$ & $(1226.2\pm8.6\pm8.8)-i(78.9\pm7.5\pm11.7)$ & $(1260-1283)-i(90-110)$ &  $(1256-1279)-i(91-107)$\\
			$f^\prime_2(1525)$ & $(1515.7\pm4.3\pm3.3)-i(33.7\pm3.8\pm4.1)$ & $(1515-1520)-i(40-46)$ & $(1488-1517)-i(34-50)$\\
			$f_2(1950)$ & $(2069.9\pm22.2\pm17.6)-i(130.0\pm26.0\pm19.3)$ & $(1830-2020)-i(110-220)$ & $(1862-2084)-i(108-269)$\\
			\hline
			\hline
		\end{tabular}
	}
	\label{tab:poles_sum}
\end{table*}

\begin{table*}[htbp]
	\centering
	\renewcommand{\arraystretch}{1.4}
	\tabcolsep=0.10cm
	\caption{The squared modulus of residues $|\mathcal{R}|^2$ and branching fractions $\mathcal{B}_{\psi(3686)}$ of $\psi(3686)\to\gamma R, R\to K_S^0 K_S^0$ for poles $R$ obtained in this work, as well as the branching fractions $\mathcal{B}_{J/\psi}$ of individual resonances $R$ in $J/\psi\to\gamma R, R\to K^0_S K^0_S$~\cite{BESIII:2018ubj}. Here, the first uncertainty is statistical and the second is systematic.}
	\scalebox{0.68}{
		\begin{tabular}{c|ccccccc}
			\hline\hline
			$|\mathcal{R}|^2$  & $f_0(1370)$ & $f_0(1500)$ & $f_0(1710)$ & $f_0(2020)$ & $f_2(1270)$ & $f_2(1525)$ & $f_2(1950)$\\ \hline
			$E_{1}$  & $19.1^{+45.6}_{-12.1}\pm{26.7}$ & $11.9^{+9.7}_{-4.9}\pm{6.8}$ & $274.2^{+40.2}_{-29.9}\pm{27.4}$ & $137.3^{+24.5}_{-22.0}\pm{37.1}$ & $7.4^{+2.0}_{-2.4}\pm{3.6}$ & $3.4^{+1.8}_{-1.6}\pm{1.1}$ & $25.5^{+17.1}_{-11.5}\pm{11.9}$\\
			$M_{2}$  & --- & --- & --- & --- & $9.3^{+2.2}_{-2.7}\pm{2.9}$ & $3.2^{+1.1}_{-1.2}\pm{1.0}$ & $5.5^{+5.5}_{-3.1}\pm{3.6}$\\
			$E_{3}$  & --- & --- & --- & --- & $3.3^{+1.9}_{-1.0}\pm{1.4}$ & $0.7^{+1.0}_{-0.4}\pm{0.4}$ & $2.6^{+3.9}_{-1.5}\pm{1.5}$\\
			Sum  & $19.1^{+45.6}_{-12.1}\pm{26.7}$ & $11.9^{+9.7}_{-4.9}\pm{6.8}$ & $274.2^{+40.2}_{-29.9}\pm{27.4}$ & $137.3^{+24.5}_{-22.0}\pm{37.1}$ & $20.1^{+3.9}_{-4.4}\pm{7.4}$ & $7.3^{+2.6}_{-2.1}\pm{2.1}$ & $33.5^{+20.7}_{-12.7}\pm{15.8}$\\ \hline\hline
			$\mathcal{B}_{\psi(3686)}$~$(\times 10^{-6})$   & $1.2^{+2.8}_{-0.7}\pm{1.6}$ & $1.7^{+1.4}_{-0.7}\pm{1.0}$ & $18.3^{+2.7}_{-2.0}\pm{1.8}$ & $4.6^{+0.8}_{-0.7}\pm{1.3}$ & $2.1^{+0.4}_{-0.5}\pm{0.8}$ & $1.4^{+0.5}_{-0.4}\pm{0.4}$ & $1.2^{+0.8}_{-0.5}\pm{0.6}$\\ 
			$\mathcal{B}_{J/\psi}$~$(\times 10^{-5})$  & $1.1^{+0.1}_{-0.1} {}^{+0.4}_{-0.3}$ & $1.6^{+0.2}_{-0.2} {}^{+0.2}_{-0.6}$ & $20.0^{+0.3}_{-0.2} {}^{+3.1}_{-1.0}$ & $27.2^{+0.8}_{-0.6} {}^{+1.7}_{-4.7}$ & $2.6^{+0.1}_{-0.1} {}^{+0.6}_{-0.2}$ & $8.0^{+0.0}_{-0.0} {}^{+0.7}_{-0.5}$ & $5.5^{+0.3}_{-0.4} {}^{+3.8}_{-1.5}$\\
			$\mathcal{B}_{\psi(3686)}/\mathcal{B}_{J/\psi}$~$(\%)$  & $11.0^{+30.6}_{-17.3}$ & $11.0^{+11.6}_{-7.8}$ & $9.2^{+1.7}_{-2.0}$ & $1.7^{+0.6}_{-0.5}$ & $8.1^{+3.4}_{-3.9}$ & $1.8^{+0.8}_{-0.7}$ & $2.3^{+1.9}_{-2.1}$\\
			\hline
			\hline
		\end{tabular}
	}
	\label{tab:poles_res_sum}
\end{table*}

\acknowledgments
The BESIII Collaboration thanks the staff of BEPCII (https://cstr.cn/31109.02.BEPC) and the IHEP computing center for their strong support. This work is supported in part by National Key R\&D Program of China under Contracts Nos. 2020YFA0406300, 2020YFA0406400, 2023YFA1606000, 2023YFA1606704; National Natural Science Foundation of China (NSFC) under Contracts Nos. 11635010, 11735014, 11935015, 11935016, 11935018, 12025502, 12035009, 12035013, 12061131003, 12192260, 12192261, 12192262, 12192263, 12192264, 12192265, 12221005, 12225509, 12235017, 12361141819; the Chinese Academy of Sciences (CAS) Large-Scale Scientific Facility Program; the CAS Center for Excellence in Particle Physics (CCEPP); Joint Large-Scale Scientific Facility Funds of the NSFC and CAS under Contract No. U1832207; CAS under Contract No. YSBR-101; 100 Talents Program of CAS; The Institute of Nuclear and Particle Physics (INPAC) and Shanghai Key Laboratory for Particle Physics and Cosmology; Agencia Nacional de Investigación y Desarrollo de Chile (ANID), Chile under Contract No. ANID PIA/APOYO AFB230003; German Research Foundation DFG under Contract No. FOR5327; Istituto Nazionale di Fisica Nucleare, Italy; Knut and Alice Wallenberg Foundation under Contracts Nos. 2021.0174, 2021.0299; Ministry of Development of Turkey under Contract No. DPT2006K-120470; National Research Foundation of Korea under Contract No. NRF-2022R1A2C1092335; National Science and Technology fund of Mongolia; National Science Research and Innovation Fund (NSRF) via the Program Management Unit for Human Resources \& Institutional Development, Research and Innovation of Thailand under Contract No. B50G670107; Polish National Science Centre under Contract No. 2019/35/O/ST2/02907; Swedish Research Council under Contract No. 2019.04595; The Swedish Foundation for International Cooperation in Research and Higher Education under Contract No. CH2018-7756; U. S. Department of Energy under Contract No. DE-FG02-05ER41374.

\bibliographystyle{JHEP}
\bibliography{reference.bib}

\providecommand{\href}[2]{#2}\begingroup\raggedright\begin{thebibliography}{10}

\bibitem{Brambilla:2005zw}
N.~Brambilla, Y.~Jia and A.~Vairo, \emph{{Model-independent study of magnetic
  dipole transitions in quarkonium}},
  \href{https://doi.org/10.1103/PhysRevD.73.054005}{\emph{Phys. Rev. D}
  {\bfseries 73} (2006) 054005}
  [\href{https://arxiv.org/abs/hep-ph/0512369}{{\ttfamily hep-ph/0512369}}].

\bibitem{Jaffe:1975fd}
R.~L. Jaffe and K.~Johnson, \emph{{Unconventional States of Confined Quarks and
  Gluons}}, \href{https://doi.org/10.1016/0370-2693(76)90423-8}{\emph{Phys.
  Lett. B} {\bfseries 60} (1976) 201}.

\bibitem{Barnes:1981kq}
T.~Barnes, F.~E. Close and S.~Monaghan, \emph{{Hyperfine Splittings of Bag
  Model Gluonia}},
  \href{https://doi.org/10.1016/0550-3213(82)90331-5}{\emph{Nucl. Phys. B}
  {\bfseries 198} (1982) 380}.

\bibitem{Barnes:1981ac}
T.~Barnes, \emph{{A Transverse Gluonium Potential Model With Breit-fermi
  Hyperfine Effects}}, \href{https://doi.org/10.1007/BF01549736}{\emph{Z. Phys.
  C} {\bfseries 10} (1981) 275}.

\bibitem{Cornwall:1982zn}
J.~M. Cornwall and A.~Soni, \emph{{Glueballs as Bound States of Massive
  Gluons}}, \href{https://doi.org/10.1016/0370-2693(83)90481-1}{\emph{Phys.
  Lett. B} {\bfseries 120} (1983) 431}.

\bibitem{Brau:2004xw}
F.~Brau and C.~Semay, \emph{{Semirelativistic potential model for glueball
  states}}, \href{https://doi.org/10.1103/PhysRevD.70.014017}{\emph{Phys. Rev.
  D} {\bfseries 70} (2004) 014017}
  [\href{https://arxiv.org/abs/hep-ph/0412173}{{\ttfamily hep-ph/0412173}}].

\bibitem{Shifman:1978by}
M.~A. Shifman, A.~I. Vainshtein and V.~I. Zakharov, \emph{{QCD and Resonance
  Physics: Applications}},
  \href{https://doi.org/10.1016/0550-3213(79)90023-3}{\emph{Nucl. Phys. B}
  {\bfseries 147} (1979) 448}.

\bibitem{Narison:1996fm}
S.~Narison, \emph{{Masses, decays and mixings of gluonia in QCD}},
  \href{https://doi.org/10.1016/S0550-3213(97)00562-2}{\emph{Nucl. Phys. B}
  {\bfseries 509} (1998) 312}
  [\href{https://arxiv.org/abs/hep-ph/9612457}{{\ttfamily hep-ph/9612457}}].

\bibitem{Bali:1993fb}
{\scshape UKQCD} collaboration, \emph{{A Comprehensive lattice study of SU(3)
  glueballs}}, \href{https://doi.org/10.1016/0370-2693(93)90948-H}{\emph{Phys.
  Lett. B} {\bfseries 309} (1993) 378}
  [\href{https://arxiv.org/abs/hep-lat/9304012}{{\ttfamily hep-lat/9304012}}].

\bibitem{Sexton:1995kd}
J.~Sexton, A.~Vaccarino and D.~Weingarten, \emph{{Numerical evidence for the
  observation of a scalar glueball}},
  \href{https://doi.org/10.1103/PhysRevLett.75.4563}{\emph{Phys. Rev. Lett.}
  {\bfseries 75} (1995) 4563}
  [\href{https://arxiv.org/abs/hep-lat/9510022}{{\ttfamily hep-lat/9510022}}].

\bibitem{Morningstar:1999rf}
C.~J. Morningstar and M.~J. Peardon, \emph{{The Glueball spectrum from an
  anisotropic lattice study}},
  \href{https://doi.org/10.1103/PhysRevD.60.034509}{\emph{Phys. Rev. D}
  {\bfseries 60} (1999) 034509}
  [\href{https://arxiv.org/abs/hep-lat/9901004}{{\ttfamily hep-lat/9901004}}].

\bibitem{Loan:2005ff}
M.~Loan, X.-Q. Luo and Z.-H. Luo, \emph{{Monte Carlo study of glueball masses
  in the Hamiltonian limit of SU(3) lattice gauge theory}},
  \href{https://doi.org/10.1142/S0217751X06029454}{\emph{Int. J. Mod. Phys. A}
  {\bfseries 21} (2006) 2905}
  [\href{https://arxiv.org/abs/hep-lat/0503038}{{\ttfamily hep-lat/0503038}}].

\bibitem{Chen:2005mg}
Y.~Chen et~al., \emph{{Glueball spectrum and matrix elements on anisotropic
  lattices}}, \href{https://doi.org/10.1103/PhysRevD.73.014516}{\emph{Phys.
  Rev. D} {\bfseries 73} (2006) 014516}
  [\href{https://arxiv.org/abs/hep-lat/0510074}{{\ttfamily hep-lat/0510074}}].

\bibitem{Richards:2010ck}
{\scshape UKQCD} collaboration, \emph{{Glueball mass measurements from improved
  staggered fermion simulations}},
  \href{https://doi.org/10.1103/PhysRevD.82.034501}{\emph{Phys. Rev. D}
  {\bfseries 82} (2010) 034501}
  [\href{https://arxiv.org/abs/1005.2473}{{\ttfamily arXiv:1005.2473}}].

\bibitem{Bali:2000vr}
{\scshape SESAM and T$\chi$L} collaboration, \emph{{Static potentials and
  glueball masses from QCD simulations with Wilson sea quarks}},
  \href{https://doi.org/10.1103/PhysRevD.62.054503}{\emph{Phys. Rev. D}
  {\bfseries 62} (2000) 054503}
  [\href{https://arxiv.org/abs/hep-lat/0003012}{{\ttfamily hep-lat/0003012}}].

\bibitem{Sun:2017ipk}
W.~Sun, L.-C. Gui, Y.~Chen, M.~Gong, C.~Liu, Y.-B. Liu et~al., \emph{{Glueball
  spectrum from $N_f=2$ lattice QCD study on anisotropic lattices}},
  \href{https://doi.org/10.1088/1674-1137/42/9/093103}{\emph{Chin. Phys. C}
  {\bfseries 42} (2018) 093103}
  [\href{https://arxiv.org/abs/1702.08174}{{\ttfamily arXiv:1702.08174}}].

\bibitem{BESIII:2015rug}
{\scshape BESIII} collaboration, \emph{{Amplitude analysis of the
  $\pi^{0}\pi^{0}$~system produced in radiative $J/\psi$~decays}},
  \href{https://doi.org/10.1103/PhysRevD.92.052003}{\emph{Phys. Rev. D}
  {\bfseries 92} (2015) 052003}
  [\href{https://arxiv.org/abs/1506.00546}{{\ttfamily arXiv:1506.00546}}],
  [Erratum: Phys.Rev.D 93, 039906 (2016)].

\bibitem{BESIII:2018ubj}
{\scshape BESIII} collaboration, \emph{{Amplitude analysis of the $K_{S}K_{S}$
  system produced in radiative $J/\psi$ decays}},
  \href{https://doi.org/10.1103/PhysRevD.98.072003}{\emph{Phys. Rev. D}
  {\bfseries 98} (2018) 072003}
  [\href{https://arxiv.org/abs/1808.06946}{{\ttfamily arXiv:1808.06946}}].

\bibitem{BESIII:2013qqz}
{\scshape BESIII} collaboration, \emph{{Partial wave analysis of $J/\psi \to
  \gamma \eta \eta$}},
  \href{https://doi.org/10.1103/PhysRevD.87.092009}{\emph{Phys. Rev. D}
  {\bfseries 87} (2013) 092009}
  [\href{https://arxiv.org/abs/1301.0053}{{\ttfamily arXiv:1301.0053}}],
  [Erratum: Phys.Rev.D 87, 119901 (2013)].

\bibitem{BESIII:2022zel}
{\scshape BESIII} collaboration, \emph{{Partial wave analysis of $J/\psi
  \rightarrow \gamma \eta' \eta'$}},
  \href{https://doi.org/10.1103/PhysRevD.105.072002}{\emph{Phys. Rev. D}
  {\bfseries 105} (2022) 072002}
  [\href{https://arxiv.org/abs/2201.09710}{{\ttfamily arXiv:2201.09710}}].

\bibitem{BESIII:2022iwi}
{\scshape BESIII} collaboration, \emph{{Partial wave analysis of $J/\psi\to
  \gamma \eta \eta^\prime$}},
  \href{https://doi.org/10.1103/PhysRevD.106.072012}{\emph{Phys. Rev. D}
  {\bfseries 106} (2022) 072012}
  [\href{https://arxiv.org/abs/2202.00623}{{\ttfamily arXiv:2202.00623}}],
  [Erratum: Phys.Rev.D 107, 079901 (2023)].

\bibitem{BESIII:2022riz}
{\scshape BESIII} collaboration.

\bibitem{Dobbs:2015dwa}
S.~Dobbs, A.~Tomaradze, T.~Xiao and K.~K. Seth, \emph{{Comprehensive Study of
  the Radiative Decays of $J/\psi$ and $\psi(2S)$ to Pseudoscalar Meson Pairs,
  and Search for Glueballs}},
  \href{https://doi.org/10.1103/PhysRevD.91.052006}{\emph{Phys. Rev. D}
  {\bfseries 91} (2015) 052006}
  [\href{https://arxiv.org/abs/1502.01686}{{\ttfamily arXiv:1502.01686}}].

\bibitem{BESIII:2024lks}
{\scshape BESIII} collaboration, \emph{{Determination of the number of
  \ensuremath{\psi}(3686) events taken at BESIII*}},
  \href{https://doi.org/10.1088/1674-1137/ad595b}{\emph{Chin. Phys. C}
  {\bfseries 48} (2024) 093001}
  [\href{https://arxiv.org/abs/2403.06766}{{\ttfamily arXiv:2403.06766}}].

\bibitem{BESIII:2009fln}
{\scshape BESIII} collaboration, \emph{{Design and Construction of the BESIII
  Detector}}, \href{https://doi.org/10.1016/j.nima.2009.12.050}{\emph{Nucl.
  Instrum. Meth. A} {\bfseries 614} (2010) 345}
  [\href{https://arxiv.org/abs/0911.4960}{{\ttfamily arXiv:0911.4960}}].

\bibitem{Yu:2016cof}
C.~Yu et~al., \emph{{BEPCII Performance and Beam Dynamics Studies on
  Luminosity}},  in \emph{{7th International Particle Accelerator Conference}},
  p.~TUYA01, 2016, \href{https://doi.org/10.18429/JACoW-IPAC2016-TUYA01}{DOI}.

\bibitem{BESIII:2020nme}
{\scshape BESIII} collaboration, \emph{{Future Physics Programme of BESIII}},
  \href{https://doi.org/10.1088/1674-1137/44/4/040001}{\emph{Chin. Phys. C}
  {\bfseries 44} (2020) 040001}
  [\href{https://arxiv.org/abs/1912.05983}{{\ttfamily arXiv:1912.05983}}].

\bibitem{Lu:2020bdc}
J.-d. Lu et~al., \emph{{Online monitoring of the center-of-mass energy from
  real data at BESIII}},
  \href{https://doi.org/10.1007/s41605-020-00188-8}{\emph{Radiat. Detect.
  Technol. Methods} {\bfseries 4} (2020) 337}.

\bibitem{Zhang:2022bdc}
J.-W. Zhang et~al., \emph{{Suppression of top-up injection backgrounds with
  offline event filter in the BESIII experiment}},
  \href{https://doi.org/10.1007/s41605-022-00331-7}{\emph{Radiat. Detect.
  Technol. Methods} {\bfseries 6} (2022) 289}.

\bibitem{Li:2017jpg}
X.~Li et~al., \emph{{Study of MRPC technology for BESIII endcap-TOF upgrade}},
  \href{https://doi.org/10.1007/s41605-017-0014-2}{\emph{Radiat. Detect.
  Technol. Methods} {\bfseries 1} (2017) 13}.

\bibitem{Guo:2017sjt}
Y.-X. Guo et~al., \emph{{The study of time calibration for upgraded end cap TOF
  of BESIII}}, \href{https://doi.org/10.1007/s41605-017-0012-4}{\emph{Radiat.
  Detect. Technol. Methods} {\bfseries 1} (2017) 15}.

\bibitem{Cao:2020ibk}
P.~Cao et~al., \emph{{Design and construction of the new BESIII endcap
  Time-of-Flight system with MRPC Technology}},
  \href{https://doi.org/10.1016/j.nima.2019.163053}{\emph{Nucl. Instrum. Meth.
  A} {\bfseries 953} (2020) 163053}.

\bibitem{GEANT4:2002zbu}
{\scshape GEANT4} collaboration, \emph{{GEANT4--a simulation toolkit}},
  \href{https://doi.org/10.1016/S0168-9002(03)01368-8}{\emph{Nucl. Instrum.
  Meth. A} {\bfseries 506} (2003) 250}.

\bibitem{Jadach:1999vf}
S.~Jadach, B.~F.~L. Ward and Z.~Was, \emph{{The Precision Monte Carlo event
  generator $KK$ for two fermion final states in $e^+e^-$ collisions}},
  \href{https://doi.org/10.1016/S0010-4655(00)00048-5}{\emph{Comput. Phys.
  Commun.} {\bfseries 130} (2000) 260}
  [\href{https://arxiv.org/abs/hep-ph/9912214}{{\ttfamily hep-ph/9912214}}].

\bibitem{Lange:2001uf}
D.~J. Lange, \emph{{The EvtGen particle decay simulation package}},
  \href{https://doi.org/10.1016/S0168-9002(01)00089-4}{\emph{Nucl. Instrum.
  Meth. A} {\bfseries 462} (2001) 152}.

\bibitem{Ping:2008zz}
R.-G. Ping, \emph{{Event generators at BESIII}},
  \href{https://doi.org/10.1088/1674-1137/32/8/001}{\emph{Chin. Phys. C}
  {\bfseries 32} (2008) 599}.

\bibitem{ParticleDataGroup:2024cfk}
{\scshape Particle Data Group} collaboration, \emph{{Review of particle
  physics}}, \href{https://doi.org/10.1103/PhysRevD.110.030001}{\emph{Phys.
  Rev. D} {\bfseries 110} (2024) 030001}.

\bibitem{Chen:2000tv}
J.~C. Chen, G.~S. Huang, X.~R. Qi, D.~H. Zhang and Y.~S. Zhu, \emph{{Event
  generator for $J/\psi$ and $\psi(2S)$ decay}},
  \href{https://doi.org/10.1103/PhysRevD.62.034003}{\emph{Phys. Rev. D}
  {\bfseries 62} (2000) 034003}.

\bibitem{Yang:2014vra}
R.-L. Yang, R.-G. Ping and H.~Chen, \emph{{Tuning and Validation of the
  Lundcharm Model with $J/\psi$ Decays}},
  \href{https://doi.org/10.1088/0256-307X/31/6/061301}{\emph{Chin. Phys. Lett.}
  {\bfseries 31} (2014) 061301}.

\bibitem{Barberio:1990ms}
E.~Barberio, B.~van Eijk and Z.~Was, \emph{{PHOTOS: A Universal Monte Carlo for
  QED radiative corrections in decays}},
  \href{https://doi.org/10.1016/0010-4655(91)90012-A}{\emph{Comput. Phys.
  Commun.} {\bfseries 66} (1991) 115}.

\bibitem{Zou:2002ar}
B.~S. Zou and D.~V. Bugg, \emph{{Covariant tensor formalism for partial wave
  analyses of psi decay to mesons}},
  \href{https://doi.org/10.1140/epja/i2002-10135-4}{\emph{Eur. Phys. J. A}
  {\bfseries 16} (2003) 537}
  [\href{https://arxiv.org/abs/hep-ph/0211457}{{\ttfamily hep-ph/0211457}}].

\bibitem{CrystalBarrel:2019zqh}
{\scshape Crystal Barrel} collaboration, \emph{{Coupled channel analysis of
  ${\bar{p}p}\,\rightarrow \,\pi ^0\pi ^0\eta $, ${\pi ^0\eta \eta }$ and
  ${K^+K^-\pi ^0}$ at 900 MeV/c and of ${\pi \pi }$-scattering data}},
  \href{https://doi.org/10.1140/epjc/s10052-020-7930-x}{\emph{Eur. Phys. J. C}
  {\bfseries 80} (2020) 453}
  [\href{https://arxiv.org/abs/1909.07091}{{\ttfamily arXiv:1909.07091}}].

\bibitem{Husken:2022yik}
N.~H\"usken, R.~E. Mitchell and E.~S. Swanson, \emph{{K-matrix analysis of
  $e^+e^-$ annihilation in the bottomonium region}},
  \href{https://doi.org/10.1103/PhysRevD.106.094013}{\emph{Phys. Rev. D}
  {\bfseries 106} (2022) 094013}
  [\href{https://arxiv.org/abs/2204.11915}{{\ttfamily arXiv:2204.11915}}].

\bibitem{James:1975dr}
F.~James and M.~Roos, \emph{{Minuit: A System for Function Minimization and
  Analysis of the Parameter Errors and Correlations}},
  \href{https://doi.org/10.1016/0010-4655(75)90039-9}{\emph{Comput. Phys.
  Commun.} {\bfseries 10} (1975) 343}.

\bibitem{Langenbruch:2019nwe}
C.~Langenbruch, \emph{{Parameter uncertainties in weighted unbinned maximum
  likelihood fits}},
  \href{https://doi.org/10.1140/epjc/s10052-022-10254-8}{\emph{Eur. Phys. J. C}
  {\bfseries 82} (2022) 393}
  [\href{https://arxiv.org/abs/1911.01303}{{\ttfamily arXiv:1911.01303}}].

\bibitem{zou:2004Angular}
S.~Dulat, B.~S. Zou and J.~M. Wu, \emph{{Angular distributions of $\psi$
  radiative decays}}, {\emph{High Energy Physics \& Nuclear Physics} {\bfseries
  28} (2004) }.

\bibitem{BESIII:2012mpj}
{\scshape BESIII} collaboration, \emph{{Search for hadronic transition
  $\chi_{cJ}\to\eta_c\pi^+\pi^-$ and observation of $\chi_{cJ}\to
  K\bar{K}\pi\pi\pi$}},
  \href{https://doi.org/10.1103/PhysRevD.87.012002}{\emph{Phys. Rev. D}
  {\bfseries 87} (2013) 012002}
  [\href{https://arxiv.org/abs/1208.4805}{{\ttfamily arXiv:1208.4805}}].

\bibitem{VonDetten:2021rax}
L.~Von~Detten, F.~No\"el, C.~Hanhart, M.~Hoferichter and B.~Kubis, \emph{{On
  the scalar $\pi K$ form factor beyond the elastic region}},
  \href{https://doi.org/10.1140/epjc/s10052-021-09169-7}{\emph{Eur. Phys. J. C}
  {\bfseries 81} (2021) 420}
  [\href{https://arxiv.org/abs/2103.01966}{{\ttfamily arXiv:2103.01966}}].

\bibitem{Rodas:2021tyb}
A.~Rodas, A.~Pilloni, M.~Albaladejo, C.~Fernandez-Ramirez, V.~Mathieu and A.~P.
  Szczepaniak, \emph{{Scalar and tensor resonances in $J/\psi $ radiative
  decays}}, \href{https://doi.org/10.1140/epjc/s10052-022-10014-8}{\emph{Eur.
  Phys. J. C} {\bfseries 82} (2022) 80}
  [\href{https://arxiv.org/abs/2110.00027}{{\ttfamily arXiv:2110.00027}}].

\end{thebibliography}\endgroup

\newpage
{\bf The BESIII Collaboration}

M.~Ablikim$^{1}$, M.~N.~Achasov$^{4,c}$, P.~Adlarson$^{77}$, X.~C.~Ai$^{82}$, R.~Aliberti$^{36}$, A.~Amoroso$^{76A,76C}$, Q.~An$^{73,59,a}$, Y.~Bai$^{58}$, O.~Bakina$^{37}$, Y.~Ban$^{47,h}$
% ...
\\
\vspace{0.2cm}
(BESIII Collaboration)\\
\vspace{0.2cm} {\it
$^{1}$ Institute of High Energy Physics, Beijing 100049, People's Republic of China\\
% ...
$^{4}$ Budker Institute of Nuclear Physics SB RAS (BINP), Novosibirsk 630090, Russia\\
% ...
$^{36}$ Johannes Gutenberg University of Mainz, Johann-Joachim-Becher-Weg 45, D-55099 Mainz, Germany\\
$^{37}$ Joint Institute for Nuclear Research, 141980 Dubna, Moscow region, Russia\\
% ...
$^{47}$ Peking University, Beijing 100871, People's Republic of China\\
% ...
$^{58}$ Southeast University, Nanjing 211100, People's Republic of China\\
$^{59}$ State Key Laboratory of Particle Detection and Electronics, Beijing 100049, Hefei 230026, People's Republic of China\\
% ...
$^{73}$ University of Science and Technology of China, Hefei 230026, People's Republic of China\\
% ...
$^{76}$ University of Turin and INFN, (A)University of Turin, I-10125, Turin, Italy; (B)University of Eastern Piedmont, I-15121, Alessandria, Italy; (C)INFN, I-10125, Turin, Italy\\
$^{77}$ Uppsala University, Box 516, SE-75120 Uppsala, Sweden\\
% ...
$^{82}$ Zhengzhou University, Zhengzhou 450001, People's Republic of China\\

\vspace{0.2cm}
$^{a}$ Deceased\\
% ...
$^{c}$ Also at the Novosibirsk State University, Novosibirsk, 630090, Russia\\
% ...
$^{h}$ Also at State Key Laboratory of Nuclear Physics and Technology, Peking University, Beijing 100871, People's Republic of China\\
% ...

}

\end{document}